# Group Divisible Codes and Their Application in the Construction of Optimal Constant-Composition Codes of Weight Three

Yeow Meng Chee, *Senior Member, IEEE*, Gennian Ge, and Alan C. H. Ling

*Abstract*—The concept of group divisible codes, a generalization of group divisible designs with constant block size, is introduced in this paper. This new class of codes is shown to be useful in recursive constructions for constant-weight and constant-composition codes. Large classes of group divisible codes are constructed which enabled the determination of the sizes of optimal constant-composition codes of weight three (and specified distance), leaving only four cases undetermined. Previously, the sizes of constant-composition codes of weight three were known only for those of sufficiently large length.

## I. INTRODUCTION

ONE generalization of constant-weight binary codes as we enlarge the alphabet from size two to beyond, is the concept of constant-composition codes. The class of constant-composition codes includes the important permutation codes and have attracted recent interest due to their numerous applications, such as in determining the zero error decision feedback capacity of discrete memoryless channels [1], multiple-access communications [2], spherical codes for modulation [3], DNA codes [4], [5], powerline communications [6], [7], and frequency hopping [8].

While constant-composition codes have been used since the early 1980s to bound error and erasure probabilities in decision feedback channels [9], their systematic study only began in late 1990s with Svanström [10]. Today, the problem of determining the maximum size of a constant-composition code constitutes a central problem in their investigation [6], [7], [11]–[20].

Our interest in this paper is in determining the maximum sizes of constant-composition codes of weight three. The techniques introduced in this paper are built upon the authors' earlier results [12], where pairwise balanced designs and group divisible designs are used to obtain optimal constant-composition codes of sufficiently large lengths. We remarked in that paper that the techniques developed therein, together with deeper methods in combinatorial design theory, can be used to derive optimal constant-composition codes of all lengths, except for a small finite set. In this paper, we show how this can be done by introducing the concept of *group divisible codes* and applying it to the problem of determining the sizes of optimal constant-composition codes of weight three. The power of group divisible codes lies in their similarity to group divisible designs, which allow the use of Wilson-type constructions [21], [22].

We begin by reviewing some coding theoretic terminology and notations.

The set of integers $\{1, 2, \ldots, n\}$ is denoted by $[n]$. The ring $\mathbb{Z}/q\mathbb{Z}$ is denoted by $\mathbb{Z}_q$, and the set of nonnegative integers and positive integers are denote by $\mathbb{Z}_{\geq 0}$ and $\mathbb{Z}_{>0}$, respectively. The notation $\lceil \cdot \rceil$ is used for multisets.

All sets considered in this paper are finite if not obviously infinite. If $X$ and $R$ are finite sets, $R^X$ denotes the set of vectors of length $|X|$, where each component of a vector $\mathsf{u} \in R^X$ has value in $R$ and is indexed by an element of $X$, that is, $\mathsf{u} = (\mathsf{u}_x)_{x \in X}$, and $\mathsf{u}_x \in R$ for each $x \in X$. A *$q$-ary code of length $n$* is a set $\mathcal{C} \subseteq \mathbb{Z}_q^X$ for some $X$ of size $n$. The elements of $\mathcal{C}$ are called *codewords*. The *Hamming norm* or the *Hamming weight* of a vector $\mathsf{u} \in \mathbb{Z}_q^X$ is defined as $\|\mathsf{u}\| = |\{x \in X : \mathsf{u}_x \neq 0\}|$. The distance induced by this norm is called the *Hamming distance*, denoted $d_H$, so that $d_H(\mathsf{u}, \mathsf{v}) = \|\mathsf{u} - \mathsf{v}\|$, for $\mathsf{u}, \mathsf{v} \in \mathbb{Z}_q^X$. The *composition* of a vector $\mathsf{u} \in \mathbb{Z}_q^X$ is the tuple $\bar{w} = [w_1, \ldots, w_{q-1}]$, where $w_j = |\{x \in X : \mathsf{u}_x = j\}|$. For any two vectors $\mathsf{u}, \mathsf{v} \in \mathbb{Z}_q^X$, define their *support* as $\operatorname{supp}(\mathsf{u}, \mathsf{v}) = \{x \in X : \mathsf{u}_x \neq \mathsf{v}_x\}$. We write $\operatorname{supp}(\mathsf{u})$ instead of $\operatorname{supp}(\mathsf{u}, 0)$ and also call $\operatorname{supp}(\mathsf{u})$ *the support of* $\mathsf{u}$.

A code $\mathcal{C}$ is said to have *distance* $d$ if $d_H(\mathsf{u}, \mathsf{v}) \geq d$ for all $\mathsf{u}, \mathsf{v} \in \mathcal{C}$. If $\|\mathsf{u}\| = w$ for every codeword $\mathsf{u} \in \mathcal{C}$, then $\mathcal{C}$ is said to be of (constant) *weight* $w$. A $q$-ary code $\mathcal{C}$ has *constant composition* $\bar{w}$ if every codeword in $\mathcal{C}$ has composition $\bar{w}$. A $q$-ary code of length $n$, distance $d$, and constant composition $\bar{w}$ is referred to as an $(n, d, \bar{w})_q$-*code*. The maximum size of an $(n, d, \bar{w})_q$-code is denoted $A_q(n, d, \bar{w})$ and the $(n, d, \bar{w})_q$-codes achieving this size are called *optimal*. Note that the following operations do not affect distance and weight properties of an $(n, d, \bar{w})_q$-code:

　i) reordering the components of $\bar{w}$;
　ii) deleting zero components of $\bar{w}$.

Manuscript received September 20, 2007; revised April 3, 2008. The work of Y. M. Chee was supported in part by the Singapore National Research Foundation, the Singapore Ministry of Education under Research Grant T206B2204 and by the Nanyang Technological University under Research Grant M58110040. The work of G. Ge was supported by in part by the National Natural Science Foundation of China under Grant 10771193, the Zhejiang Provincial Natural Science Foundation of China, and the Program for New Century Excellent Talents in University. This work was done when A. C. H. Ling was on sabbatical leave at the Division of Mathematical Sciences, School of Physical and Mathematical Sciences, Nanyang Technological University, Singapore.

Y. M. Chee is with the Division of Mathematical Sciences, School of Physical and Mathematical Sciences, Nanyang Technological University, Singapore 637371, Singapore (e-mail: ymchee@ntu.edu.sg).

G. Ge is with the Department of Mathematics, Zhejiang University, Hangzhou 310027, Zhejiang, China (e-mail: gnge@zju.edu.cn).

A. C. H. Ling is with the Department of Computer Science, University of Vermont, Burlington, VT 05405 USA (e-mail: aling@emba.uvm.edu).

Communicated by T. Etzion, Associate Editor for Coding Theory.

Digital Object Identifier 10.1109/TIT.2008.926349

Consequently, throughout this paper, we restrict our attention to those compositions $\bar{w} = [w_1, \ldots, w_{q-1}]$, where $w_1 \geq \cdots \geq w_{q-1} \geq 1$.

Suppose $\mathsf{u} \in \mathbb{Z}_q^X$ is a codeword of an $(n, d, \bar{w})_q$-code, where $\bar{w} = [w_1, \ldots, w_{q-1}]$. Let $w = \sum_{i=1}^{q-1} w_i$. We can represent $\mathsf{u}$ equivalently as a $w$-tuple $\langle a_1, a_2, \ldots, a_w \rangle \in X^w$, where

$$\mathsf{u}_{a_1} = \cdots = \mathsf{u}_{a_{w_1}} = 1$$
$$\mathsf{u}_{a_{w_1+1}} = \cdots = \mathsf{u}_{a_{w_1+w_2}} = 2$$
$$\vdots$$
$$\mathsf{u}_{a_{\sum_{i=1}^{q-2} w_i + 1}} = \cdots = \mathsf{u}_{a_w} = q - 1.$$

Throughout this paper, we shall often represent codewords of constant-composition codes in this form. This has the advantage of being more succinct and more flexible in manipulation.

Since the distance between any two distinct codewords of a constant-composition code $\mathcal{C}$ of weight $w$ is at least two and at most $2w$, and that $\mathrm{supp}(\mathsf{u}), \mathsf{u} \in \mathcal{C}$, are pairwise disjoint if $\mathcal{C}$ has distance $2w$, we have

*Proposition 1.1:*

$$A_q(n, d, [w_1, \ldots, w_{q-1}]) = \begin{cases} \binom{n}{\sum_{i=1}^{q-1} w_i} \binom{\sum_{i=1}^{q-1} w_i}{w_1, \ldots, w_{q-1}}, & \text{if } d \leq 2 \\ \left\lfloor \frac{n}{\sum_{i=1}^{q-1} w_i} \right\rfloor, & \text{if } d = 2\sum_{i=1}^{q-1} w_i \\ 1, & \text{if } d \geq 2\sum_{i=1}^{q-1} w_i + 1. \end{cases}$$

*Proof:* Let $\bar{w} = [w_1, \ldots, w_{q-1}]$. When $d \leq 2$, the optimal $(n, d, \bar{w})_q$-code contains all vectors with composition $\bar{w}$ as codewords. When $d = 2\sum_{i=1}^{q-1} w_i$, all codewords must have disjoint supports. No pair of codewords in a $(n, d, \bar{w})_q$-code can be distance $2\sum_{i=1}^{q-1} w_i + 1$ apart. $\square$

Henceforth, we need only concern ourselves with $d \in \{3, 4, 5\}$ when we study constant-composition codes of weight three.

## II. STATE OF AFFAIRS

Constant-composition codes of weight three can be classified as follows:
  i) $q = 2$ and $\bar{w} = [3]$;
  ii) $q = 3$ and $\bar{w} = [2, 1]$;
  iii) $q = 4$ and $\bar{w} = [1, 1, 1]$.

The value of $A_2(n, d, [3])$ is a classical result in binary constant-weight codes and is given below (note that $A_2(n, d, [w]) = A_2(n, d+1, [w])$ when $d$ is odd, and so the value of $A_2(n, 5, [3]) = A_2(n, 6, [3])$, which can be obtained from Proposition 1.1, is omitted).

*Theorem 2.1 (Schönheim [23], Spencer [24]):*

$$A_2(n, d, [3]) = \begin{cases} \left\lfloor \frac{n}{3} \left\lfloor \frac{n-1}{2} \right\rfloor \right\rfloor, & \text{if } d \in \{3, 4\} \text{ and } n \not\equiv 5 \pmod{6} \\ \left\lfloor \frac{n}{3} \left\lfloor \frac{n-1}{2} \right\rfloor \right\rfloor - 1, & \text{if } d \in \{3, 4\} \text{ and } n \equiv 5 \pmod{6}. \end{cases}$$

The value of $A_3(n, d, [2, 1])$ was investigated by Svanström [10], [19] and Chee et al. [12]. Svanström [10], [19] determined that $A_3(n, 5, [2, 1]) = \lfloor n/2 \rfloor$ and $A_3(n, 3, [2, 1]) = n\lfloor(n-1)/2\rfloor$ for all $n$, and determined that $A_3(n, 4, [2, 1]) = n(n-2)/4$ if $n$ is even. In trying to determine $A_3(n, 4, [2, 1])$ for $n$ odd, the authors recently discovered a result based on the closure of pairwise balanced designs that enables the determination of $A_q(n, d, \bar{w})$ for all sufficiently large $n$, from just a single example of an optimal $(n, d, \bar{w})_q$-code, provided $d$ obeys a certain bound [12]. Using this technique, the following was shown:
  i) $A_3(n, 4, [2, 1]) = n(n-1)/4$ for all $n \equiv 1 \pmod{4}$;
  ii) $A_3(n, 4, [2, 1]) = (n-1)^2/4 + \lfloor (n-3)/12 \rfloor$ for all sufficiently large $n \equiv 3 \pmod{4}$;
  iii) $A_4(n, 3, [1, 1, 1]) = n(n-1)$ for all $n \geq 4$, except for $n \in \{5, 6\}$ and except possibly for $n \in \{44, 47, 51, 54, 59, 62, 158, 167, 173\}$;
  iv) $A_4(n, 4, [1, 1, 1]) = n\lfloor (n-1)/2 \rfloor$ for all sufficiently large $n$;
  v) $A_4(n, 5, [1, 1, 1]) = n$, for all $n \geq 7$.

The purpose of this paper is to determine the following:
  i) $A_3(n, 4, [2, 1])$ for all $n \equiv 3 \pmod{4}$;
  ii) $A_4(n, 3, [1, 1, 1])$ for
  $$n \in \{44, 47, 51, 54, 59, 62, 158, 167, 173\};$$
  iii) $A_4(n, 4, [1, 1, 1])$ for all $n \notin \{9, 13, 15, 17\}$;
thereby completing the determination of the sizes of optimal constant-composition codes of weight three, except for four cases.

Let

$$U(n, 4, [2, 1]) = \frac{(n-1)^2}{4} + \left\lfloor \frac{n-3}{12} \right\rfloor$$
$$U(n, 3, [1, 1, 1]) = n(n-1)$$
$$U(n, 4, [1, 1, 1]) = n \left\lfloor \frac{n-1}{2} \right\rfloor.$$

The bounds

$$A_3(n, 4, [2, 1]) \leq U(n, 4, [2, 1]) \quad \text{when } n \equiv 3 \pmod{4}$$
$$A_4(n, 3, [1, 1, 1]) \leq U(n, 3, [1, 1, 1])$$
$$A_4(n, 4, [1, 1, 1]) \leq U(n, 4, [1, 1, 1])$$

have already been established previously [12], [19], so we focus on the construction of constant-composition codes meeting these upper bounds.

## III. GROUP DIVISIBLE DESIGNS AND GROUP DIVISIBLE CODES

Central to our construction are the notions of *group divisible designs* and a generalization that we call *group divisible codes*. We begin by defining them.

### A. Group Divisible Designs

A *set system* is a pair $(X, \mathcal{A})$, where $X$ is a finite set of *points* and $\mathcal{A} \subseteq 2^X$, whose elements are called *blocks*. The *order* of the set system is $|X|$, the number of points. For a set of nonnegative integers $K$, a set system $(X, \mathcal{A})$ is said to be $K$-*uniform* if $|A| \in K$ for all $A \in \mathcal{A}$.

Let $(X, \mathcal{A})$ be a set system and $\mathcal{G} = \{G_1, \ldots, G_t\}$ be a partition of $X$ into subsets, called *groups*. The triple $(X, \mathcal{G}, \mathcal{A})$ is a *group divisible design* (GDD) when every 2-subset of $X$ not contained in a group appears in exactly one block and $|A \cap G| \leq 1$ for all $A \in \mathcal{A}$ and $G \in \mathcal{G}$. We denote a GDD $(X, \mathcal{G}, \mathcal{A})$ by $K$-GDD if $(X, \mathcal{A})$ is $K$-uniform. The *type* of a GDD $(X, \mathcal{G}, \mathcal{A})$ is the multiset $\lceil |G| : G \in \mathcal{G} \rceil$. We use the exponential notation to describe the type of a GDD: a GDD of type $g_1^{t_1} \cdots g_s^{t_s}$ is a GDD where there are exactly $t_i$ groups of size $g_i, 1 \leq i \leq s$.

A *parallel class* in a GDD $(X, \mathcal{G}, \mathcal{A})$ is a subset $\mathcal{A}' \subseteq \mathcal{A}$ such that each point $x \in X$ is contained in exactly one block in $\mathcal{A}'$, and a *holey parallel class* is a subset $\mathcal{A}' \subseteq \mathcal{A}$ such that for some $G \in \mathcal{G}$, each point $x \in X \setminus G$ is contained in exactly one block in $\mathcal{A}'$, and no point of $G$ is contained in any block in $\mathcal{A}'$; in other words, $\mathcal{A}'$ is a partition of $X \setminus G$. A *resolvable* GDD (RGDD) is a GDD $(X, \mathcal{G}, \mathcal{A})$ in which $\mathcal{A}$ can be partitioned into parallel classes, and a $K$-*frame* is a $K$-GDD $(X, \mathcal{G}, \mathcal{A})$ in which $\mathcal{A}$ can be partitioned into holey parallel classes. In particular, a $\{3\}$-frame is called a *Kirkman frame*. We have the following known results on the existence of RGDD's and frames.

*Theorem 3.1 (Rees and Stinson [25], Rees [26]):* There exists a $\{3\}$-RGDD of type $g^t$ if and only if $t \geq 3, gt \equiv 0 \pmod{3}$, and $g(t-1) \equiv 0 \pmod{2}$, except when $(g, t) \in \{(2, 3), (2, 6), (6, 3)\}$.

*Theorem 3.2 (Stinson [27]):* A Kirkman frame of type $g^t$ exists if and only if $t \geq 4, g \equiv 0 \pmod{2}$, and $g(t-1) \equiv 0 \pmod{3}$.

*Theorem 3.3 (Adding y Points to a Frame):* Let $y \in \mathbb{Z}_{\geq 0}$. Suppose there exists a $K$-frame $(X, \mathcal{G}, \mathcal{A})$ of type $g^t$. Then there exists a $K'$-GDD of type $g^{t-1}(g+y)^1$, where $K' = \{k, k+1 : k \in K\}$.

*Proof:* Let $\infty_1, \ldots, \infty_y \notin X, \mathcal{B}_i \subseteq \mathcal{A}$ for $1 \leq i \leq y$ be holey parallel classes missing a picked group $G \in \mathcal{G}$. Then $(X', \mathcal{G}', \mathcal{A}')$, where

$$X' = X \cup \{\infty_1, \ldots, \infty_y\}$$
$$\mathcal{G}' = (\mathcal{G} \setminus \{G\}) \cup \{G \cup \{\infty_1, \ldots, \infty_y\}\}$$
$$\mathcal{A}' = (\mathcal{A} \setminus (\cup_{i=1}^y \mathcal{B}_i)) \cup (\cup_{i=1}^y \{B \cup \{\infty_i\} : B \in \mathcal{B}_i\})$$

is a $K'$-GDD of type $g^{t-1}(g+y)^1$.

A *Latin square of side m* is an $m \times m$ array in which each cell contains a single element from a *symbol set S* of cardinality $m$, such that each element of $S$ appears exactly once in each row, and exactly once in each column. A *transversal design* $\mathrm{TD}(k, m)$ is a $\{k\}$-GDD of type $m^k$. A Latin square of side $m$ is equivalent to a $\mathrm{TD}(3, m)$. The following result on the existence of transversal designs (see [28]) is used without explicit reference throughout the paper.

*Theorem 3.4:* Let $\mathrm{TD}(k)$ denote the set of positive integers $m$ such that there exists a $\mathrm{TD}(k, m)$. Then, we have
  i) $\mathrm{TD}(3) = \mathbb{Z}_{>0}$;
  ii) $\mathrm{TD}(4) = \mathbb{Z}_{>0} \setminus \{2, 6\}$;
  iii) $\mathrm{TD}(5) = \mathbb{Z}_{>0} \setminus \{2, 3, 6, 10\}$;
  iv) $\mathrm{TD}(6) = \mathbb{Z}_{>0} \setminus \{2, 3, 4, 6, 10, 14, 18, 22\}$.

GDDs of different types can be obtained from transversal designs by truncating groups (Hanani [29]) or truncating blocks.

*Theorem 3.5 (Truncating Groups):* Let $k$ be an integer, $k \geq 2$. Let $K = \{k, k+1, \ldots, k+s\}$. Suppose that there exists a $\mathrm{TD}(k+s, m)$. Let $g_1, g_2, \ldots, g_s$ be integers satisfying $0 \leq g_i \leq m, 1 \leq i \leq s$. Then there exists a $K$-GDD of type $m^k g_1^1 g_2^1 \cdots g_s^1$.

*Theorem 3.6 (Truncating Blocks):* Let $k$ be an integer, $k \geq 2$, and let $0 \leq s \leq k$. Suppose there exists a $\mathrm{TD}(k, m)$. Then there exists a $\{k-s, k-1, k\}$-GDD of type $(m-1)^s m^{k-s}$.

*Proof:* Delete $s$ points lying in the same block from a $\mathrm{TD}(k, m)$. □

Another useful notion is that of an *incomplete transversal design*. An *incomplete transversal design of group size $n$, block size $k$, and hole size $h$*, denoted $\mathrm{ITD}(k, n; h)$ is a quadruple $(X, \mathcal{G}, H, \mathcal{A})$ such that
  i) $(X, \mathcal{A})$ is a $\{k\}$-uniform set system of order $kn$;
  ii) $\mathcal{G}$ is a partition of $X$ into subsets, called *groups*, each of size $n$;
  iii) $H \subseteq X$, with $|G \cap H| = h$ for all $G \in \mathcal{G}$;
  iv) every 2-subset of $X$ is either
     - contained in $H$ and not contained in any blocks of $\mathcal{A}$;
     - contained in a group and not contained in any blocks of $\mathcal{A}$; or
     - contained in neither $H$ nor a group, and contained in exactly one block of $\mathcal{A}$.

*Theorem 3.7 (Heinrich and Zhu [30]):* For $n > h > 0$, an $\mathrm{ITD}(4, n; h)$ exists if and only if $n \geq 3h, (n, h) \neq (6, 1)$.

### B. Group Divisible Codes

Given $\mathsf{u} \in \mathbb{Z}_q^X$ and $Y \subseteq X$, the *restriction of* $\mathsf{u}$ *to* $Y$, written $\mathsf{u}|_Y$, is the vector $\mathsf{v} \in \mathbb{Z}_q^X$ such that

$$\mathsf{v}_x = \begin{cases} \mathsf{u}_x, & \text{if } x \in Y \\ 0, & \text{if } x \in X \setminus Y. \end{cases}$$

The *constriction of* $\mathsf{u}$ *to* $Y$, written $\mathsf{u}|^Y$, is the vector $\mathsf{v} \in \mathbb{Z}_q^Y$ such that $\mathsf{v} = (\mathsf{u}_x)_{x \in Y}$.

A *group divisible code* (GDC) of distance $d$ is a triple $(X, \mathcal{G}, \mathcal{C})$, where $\mathcal{G} = \{G_1, \ldots, G_t\}$ is a partition of $X$ with cardinality $|X| = n$ and $\mathcal{C} \subseteq \mathbb{Z}_q^X$ is a $q$-ary code of length $n$, such that $d_H(\mathsf{u}, \mathsf{v}) \geq d$ for all distinct $\mathsf{u}, \mathsf{v} \in \mathcal{C}$, and $\|\mathsf{u}|_{G_i}\| \leq 1$ for all $\mathsf{u} \in \mathcal{C}, 1 \leq i \leq t$. Elements of $\mathcal{G}$ are called *groups*. We denote a GDC $(X, \mathcal{G}, \mathcal{C})$ of distance $d$ as $w$-GDC$(d)$ if $\mathcal{C}$ is of constant weight $w$. If we want to emphasize the composition of the codewords, we denote the GDC as $\bar{w}$-GDC$(d)$ when every $\mathsf{u} \in \mathcal{C}$ has composition $\bar{w}$. The *type* of a GDC $(X, \mathcal{G}, \mathcal{C})$ is the multiset $\lceil |G| : G \in \mathcal{G} \rceil$. As in the case of GDD's, the exponential notation is used to describe the type of a GDC. The *size* of a GDC $(X, \mathcal{G}, \mathcal{C})$ is $|\mathcal{C}|$.

Note that an $(n, d, \bar{w})_q$-code of size $s$ is equivalent to a $\bar{w}$-GDC$(d)$ of type $1^n$ and size $s$.

*Example 3.1:* Let $X = \mathbb{Z}_6, \mathcal{G} = \{\{0, 3\} + i : i \in \mathbb{Z}_3\}$, and $\mathcal{C}$ be the set of all cyclic shifts of the vector 210001 $\in \mathbb{Z}_3^X$. Then $(X, \mathcal{G}, \mathcal{C})$ is a $[2, 1]$-GDC$(4)$ of type $2^3$, and $\mathcal{C}$ is an optimal $(6, 4, [2, 1])_3$-code of size six.

*Example 3.2:* Let $X = \mathbb{Z}_{15}, \mathcal{G} = \{\{0, 5, 10\} + i : i \in \mathbb{Z}_5\}$, and $\mathcal{C}$ be the set of all cyclic shifts of the vectors 110200000000000, 100010000000200, 100000100000020 $\in$

```
Input:      (master) GDD D = (X, G, A);
            weight function ω : X → Z_{≥0};
            (ingredient) K-GDD D_A = (X_A, G_A, B_A) of type ⌈ω(a) : a ∈ A⌉ for each block A ∈ A, where
                X_A = ∪_{a∈A}{{a} × {1,...,ω(a)}}, and
                G_A = {{a} × {1,...,ω(a)} : a ∈ A}.
Output:     K-GDD D* = (X*, G*, A*) of type ⌈∑_{x∈G} ω(x) : G ∈ G⌉, where
                X* = ∪_{x∈X}({x} × {1,...,ω(x)}),
                G* = {∪_{x∈G}({x} × {1,...,ω(x)}) : G ∈ G}, and
                A* = ∪_{A∈A} B_A.
Notation:   D* = WFC(D, ω, {D_A : A ∈ A}).
Note:       By convention, for x ∈ X, {x} × {1,...,ω(x)} = ∅ if ω(x) = 0.
```

Fig. 1. Wilson's Fundamental Construction for GDDs.

$\mathbb{Z}_3^X$. Then $(X, \mathcal{G}, \mathcal{C})$ is a $[2,1]$-GDC(4) of type $3^5$, and $\mathcal{C}$ is a $(15, 4, [2,1])_3$-code of size 45.

Often, constant-composition codes of larger size can be obtained from GDCs via the following simple observation.

*Proposition 3.1 (Filling in Groups):* Let $d \leq 2(w-1)$. Suppose there exists a $w$-GDC($d$) $(X, \mathcal{G}, \mathcal{C})$ of type $g_1^{t_1} \cdots g_s^{t_s}$ and size $a$. Suppose further that for each $i, 1 \leq i \leq s$, there exists a $(g_i, d, w)_q$-code $\mathcal{C}_i$ of size $b_i$, then there exists a $(\sum_{i=1}^{s} t_i g_i, d, w)_q$-code $\mathcal{C}'$ of size $a + \sum_{i=1}^{s} t_i b_i$. In particular, if $\mathcal{C}$ and $\mathcal{C}_i, 1 \leq i \leq s$, are of constant composition $\bar{w}$, then $\mathcal{C}'$ is also of constant composition $\bar{w}$.

*Proof:* Let $(X, \mathcal{G}, \mathcal{C})$ be a $w$-GDC($d$) of type $g_1^{t_1} \cdots g_s^{t_s}$ and let $n = \sum_{i=1}^{s} t_i g_i$. For each $G \in \mathcal{G}$, we put a $(|G|, d, w)_q$-code on $G$. Now, the distance between any two codewords from codes on distinct groups is $2w$, and the distance between any two codewords, one from $\mathcal{C}$ and one from a code on a group, is at least $2(w-1)$. Since $d \leq 2(w-1)$, the resulting code is an $(n, d, w)_q$-code. □

*Example 3.3:* Filling in the groups of the $[2,1]$-GDC(4) of type $3^5$ in Example 3.2 with a trivial $(3, 4, [2,1])_3$-code of size one gives a $(15, 4, [2,1])_3$-code of size $45 + 5 = 50$. This constant-composition code is optimal.

There is an obvious generalization of Proposition 3.1 to allow filling in of only some of the groups, not necessarily all the groups. The example below illustrates this.

*Example 3.4:* Filling in four of the five groups of the $[2,1]$-GDC(4) of type $3^5$ in Example 3.2, with a trivial $(3, 4, [2,1])_3$-code of size one, gives a $[2,1]$-GDC(4) of type $1^{12}3^1$ having size $45 + 4 = 49$.

The following is another useful construction for constant-composition codes from GDC's.

*Proposition 3.2 (Adjoining y Points):* Let $y \in \mathbb{Z}_{\geq 0}$. Suppose there exists a (master) $w$-GDC($d$) of type $g_1^{t_1} \cdots g_s^{t_s}$ and size $a$, and suppose the following (ingredients) also exist:
  i) a $(g_1 + y, d, w)_q$-code of size $b$;
  ii) a $w$-GDC($d$) of type $1^{g_i} y^1$ and size $c_i$ for $2 \leq i \leq s$;
  iii) a $w$-GDC($d$) of type $1^{g_1} y^1$ and size $c_1$ if $t_1 \geq 2$.
Then, there exists a $(y + \sum_{i=1}^{s} t_i g_i, d, w)_q$-code of size

$$a + b + (t_1 - 1)c_1 + \sum_{i=2}^{s} t_i c_i.$$

Furthermore, if the master and ingredient codes are of constant composition, then so is the resulting code.

*Proof:* Let $(X, \mathcal{G}, \mathcal{C})$ be a $w$-GDC($d$) with $\mathcal{G} = \{G_1, \ldots, G_m\}$, and let $Y$ be a set of size $y$ disjoint from $X$. Let $\mathcal{C}'$ be a $(|G_1| + y, d, w)_q$-code and let $(G_i \cup Y, \{\{x\} : x \in G_i\} \cup \{Y\}, \mathcal{C}_i)$ be a $w$-GDC($d$) for each $i \in \{2, \ldots, m\}$. Then

$$\mathcal{C}^* = \mathcal{C} \cup \mathcal{C}' \cup (\cup_{i=2}^{m} \mathcal{C}_i)$$

is the required $(|X| + |Y|, d, w)_q$-code of size $|C| + |C'| + \sum_{i=2}^{m} |C_i|$. □

To apply Propositions 3.1 and 3.2, we require the existence of large classes of GDCs. The next theorem is a direct analogue of Wilson's Fundamental Construction for GDDs [21] (shown in Fig. 1), applied to GDCs.

*Theorem 3.8 (Fundamental Construction):* Let $d \leq 2(w-1)$, $\mathcal{D} = (X, \mathcal{G}, \mathcal{A})$ be a (master) GDD, and $\omega : X \to \mathbb{Z}_{\geq 0}$ be a weight function. Suppose that for each $A \in \mathcal{A}$, there exists an (ingredient) $w$-GDC($d$) of type $\lceil \omega(a) : a \in A \rceil$. Then there exists a $w$-GDC($d$) $\mathcal{D}^*$ of type $\lceil \sum_{x \in G} \omega(x) : G \in \mathcal{G} \rceil$. Furthermore, if the ingredient GDC's are of constant composition $\bar{w}$, then $\mathcal{D}^*$ is also of constant composition $\bar{w}$.

*Proof:* For each $A \in \mathcal{A}$, let $\mathcal{D}_A = (X_A, \mathcal{G}_A, \mathcal{C}_A)$ be a $\bar{w}$-GDC($d$) of type $\lceil \omega(a) : a \in A \rceil$, where

$$X_A = \bigcup_{a \in A} \{\{a\} \times \{1, \ldots, \omega(a)\}\}$$
$$\mathcal{G}_A = \{\{a\} \times \{1, \ldots, \omega(a)\} : a \in A\}.$$

Then $(X^*, \mathcal{G}^*, \mathcal{C}^*)$ is a $\bar{w}$-GDC($d$) of type $\lceil \sum_{x \in G} \omega(x) : G \in \mathcal{G} \rceil$, where

$$X^* = \bigcup_{x \in X} (\{x\} \times \{1, \ldots, \omega(x)\})$$
$$\mathcal{G}^* = \left\{ \bigcup_{x \in G} (\{x\} \times \{1, \ldots, \omega(x)\}) : G \in \mathcal{G} \right\}$$
$$\mathcal{C}^* = \bigcup_{A \in \mathcal{A}} \mathcal{C}_A.$$

If in the Fundamental Construction, $\omega(x) = c$ for all $x \in X$, the construction is also known as *inflating the master GDD by $c$*.

The following results provide large classes of GDDs for the constructions described above.

TABLE I
GROUPS AND PRESTRUCTURES OF SOME $\{3,4\}$-GDDs. NOTE THAT $\mathbb{Z}_{[a,b]}$ DENOTES THE SET $\{x \in \mathbb{Z} : a \leq x \leq b\}$

| Type | Groups | Prestructure |
|---|---|---|
| $5^3 6^1$ | $\mathbb{Z}_{[0,4]}, \mathbb{Z}_{[5,9]}, \mathbb{Z}_{[10,14]}, \mathbb{Z}_{[15,20]}$ | $\{0,7,11,15\}$ $\{0,6,12,16\}$ $\{1,5,12,17\}$ $\{1,7,10,19\}$ $\{2,5,11,18\}$ $\{2,6,10,20\}$ |
| $5^6 4^1$ | $\mathbb{Z}_{[0,4]}, \mathbb{Z}_{[5,9]}, \mathbb{Z}_{[10,14]}, \mathbb{Z}_{[15,19]},$ $\mathbb{Z}_{[20,24]}, \mathbb{Z}_{[25,29]}, \mathbb{Z}_{[30,33]}$ | $\{0,5,10,30\}$ $\{15,20,25,30\}$ $\{1,6,11,30\}$ $\{16,21,26,30\}$ $\{2,7,12,30\}$ $\{17,22,27,30\}$ $\{3,8,13,30\}$ $\{18,23,28,30\}$ $\{4,9,14,30\}$ $\{19,24,29,30\}$ |
| $8^4 9^1$ | $\mathbb{Z}_{[0,8]}, \mathbb{Z}_{[10,17]}, \mathbb{Z}_{[18,25]}, \mathbb{Z}_{[26,33]}, \mathbb{Z}_{[34,40]} \cup \{9\}$ | $\{10,18,26,34\}$ $\{11,19,27,35\}$ $\{12,20,28,36\}$ $\{13,21,29,37\}$ $\{14,22,30,38\}$ $\{15,23,31,39\}$ $\{16,24,32,40\}$ $\{9,17,25,33\}$ |
| $9^3 14^1$ | $\mathbb{Z}_{[0,8]}, \mathbb{Z}_{[9,17]}, \mathbb{Z}_{[18,26]}, \mathbb{Z}_{[27,40]}$ | $\{0,9,18,27\}$ $\{1,10,19,28\}$ $\{2,11,20,29\}$ $\{3,12,21,30\}$ $\{4,13,22,31\}$ $\{5,14,23,32\}$ $\{6,15,24,33\}$ $\{0,10,20,34\}$ $\{1,11,21,35\}$ $\{2,12,22,36\}$ $\{3,13,23,37\}$ $\{4,14,24,38\}$ $\{5,15,18,39\}$ $\{6,9,19,40\}$ |
| $9^3 15^1 2^1$ | $\mathbb{Z}_{[0,14]}, \mathbb{Z}_{[15,23]}, \mathbb{Z}_{[24,32]}, \mathbb{Z}_{[33,41]}, \mathbb{Z}_{[42,43]}$ | $\{0,15,24,33\}$ $\{1,16,25,34\}$ $\{2,17,26,35\}$ $\{3,18,27,36\}$ $\{4,19,28,37\}$ $\{5,20,29,39\}$ $\{6,21,30,39\}$ $\{7,22,31,40\}$ $\{8,23,32,41\}$ $\{9,15,25,35\}$ $\{10,16,26,33\}$ $\{11,17,24,34\}$ $\{12,15,26,34\}$ $\{13,16,24,35\}$ $\{14,17,25,33\}$ |
| $9^4 14^1$ | $\mathbb{Z}_{[0,8]}, \mathbb{Z}_{[9,17]}, \mathbb{Z}_{[18,26]}, \mathbb{Z}_{[27,40]}, \mathbb{Z}_{[41,49]}$ | $\{0,9,18,41\}$ $\{1,10,19,42\}$ $\{2,11,20,43\}$ $\{3,12,21,44\}$ $\{4,13,22,45\}$ $\{5,14,23,46\}$ $\{6,15,24,47\}$ $\{7,16,25,48\}$ $\{8,17,26,49\}$ |
| $7^6 9^1 8^1$ | $\mathbb{Z}_{[0,8]}, \mathbb{Z}_{[9,15]}, \mathbb{Z}_{[16,22]}, \mathbb{Z}_{[23,29]},$ $\mathbb{Z}_{[30,36]}, \mathbb{Z}_{[37,43]}, \mathbb{Z}_{[44,50]}, \mathbb{Z}_{[51,58]}$ | $\{9,16,23,51\}$ $\{10,17,24,52\}$ $\{11,18,25,53\}$ $\{12,19,26,54\}$ $\{9,17,25,55\}$ $\{10,18,26,56\}$ $\{11,19,23,57\}$ $\{12,16,24,58\}$ |

*Theorem 3.9 (Colbourn, Hoffman, and Rees [31]):* Let $g, t, u \in \mathbb{Z}_{\geq 0}$. There exists a $\{3\}$-GDD of type $g^t u^1$ if and only if the following conditions are all satisfied:
   i) if $g > 0$ then $t \geq 3$, or $t = 2$ and $u = g$, or $t = 1$ and $u = 0$, or $t = 0$;
   ii) $u \leq g(t-1)$ or $gt = 0$;
   iii) $g(t-1) + u \equiv 0 \pmod{2}$ or $gt = 0$;
   iv) $gt \equiv 0 \pmod{2}$ or $u = 0$;
   v) $g^2 \binom{t}{2} + gtu \equiv 0 \pmod{3}$.

*Theorem 3.10 (Brouwer, Schrijver, and Hanani [32]):* There exists a $\{4\}$-GDD of type $g^t$ if and only if $t \geq 4$ and
   i) $g \equiv 1$ or $5 \pmod{6}$ and $t \equiv 1$ or $4 \pmod{12}$; or
   ii) $g \equiv 2$ or $4 \pmod{6}$ and $t \equiv 1 \pmod{3}$; or
   iii) $g \equiv 3 \pmod{6}$ and $t \equiv 0$ or $1 \pmod{4}$; or
   iv) $g \equiv 0 \pmod{6}$,
with the two exceptions of types $2^4$ and $6^4$, for which $\{4\}$-GDD's do not exist.

*Theorem 3.11 (Rees and Stinson [33]):* A $\{4\}$-GDD of type $3^t u^1$ exists if and only if either
   i) $t \equiv 0 \pmod 4$ and $u \equiv 0 \pmod 3, 0 \leq u \leq (3t-6)/2$; or
   ii) $t \equiv 1 \pmod 4$ and $u \equiv 0 \pmod 6, 0 \leq u \leq (3t-3)/2$; or
   iii) $t \equiv 3 \pmod 4$ and $u \equiv 3 \pmod 6, 0 < m \leq (3t-3)/2$.

*Theorem 3.12 (Ge and Rees [34]):* There exists a $\{4\}$-GDD of type $6^t u^1$ for every $t \geq 4$ and $u \equiv 0 \pmod 3$ with $0 \leq u \leq 3t - 3$, except for $(t, u) = (4, 0)$ and except possibly for $(t, u) \in \{(7, 15), (11, 21), (11, 24), (11, 27), (13, 27), (13, 33), (17, 39), (17, 42), (19, 45), (19, 48), (19, 51), (23, 60), (23, 63)\}$.

## IV. SOME SMALL GDDs, GDCs, AND OPTIMAL CODES

In this section, we present some small GDDs, GDCs, and optimal constant-composition codes, whose existence is required in establishing subsequent results.

### A. Some $\{3,4\}$-GDDs

To construct small $\{3,4\}$-GDD's, we use the hill-climbing algorithm described in [35]. Suppose $(X, \mathcal{G}, \mathcal{A})$ is a $\{3,4\}$-GDD. We call the set $\{A \in \mathcal{A} : |A| = 4\}$ the *prestructure* of the GDD. On given a prestructure, the hill-climbing algorithm quickly finds a set of blocks of size three that can be added to complete it to a $\{3,4\}$-GDD of a specified type.

*Proposition 4.1:* There exist $\{3,4\}$-GDDs of the following types:
   i) $5^3 6^1$;
   ii) $5^6 4^1$;
   iii) $8^4 9^1$;
   iv) $9^3 14^1$;
   v) $9^3 15^1 2^1$;
   vi) $9^4 14^1$;
   vii) $7^6 9^1 8^1$.

   *Proof:* The groups and prestructures for $\{3,4\}$-GDDs of the required types are provided in Table I. We omit listing the blocks of size three since they exhibit no particular structure, are space-consuming, and can be found easily and quickly with a hill-climbing algorithm. We refer the interested reader to the first author's website at ⟨https://www1.spms.ntu.edu.sg/ym-chee/34gdd.php⟩ for a complete description of these GDDs.

### B. Some Optimal $(n, 4, [2, 1])_3$-Codes

*Proposition 4.2:* $A_3(n, 4, [2, 1]) = U(n, 4, [2, 1])$ for all $n \equiv 3 \pmod 4, 3 \leq n \leq 35$.

   *Proof:* $A_3(n, 4, [2, 1]) = U(n, 4, [2, 1])$ has been shown to hold for $n \in \{7, 11\}$ by Svanström [19] and for all $n \equiv 3 \pmod 4, 15 \leq n \leq 31$ by Chee et al. [12]. An optimal $(35, 4, [2, 1])_3$-code is given in Table II.

### C. Some $[1, 1, 1]$-GDC (3)

*Proposition 4.3:* There exists a $[1, 1, 1]$-GDC (3) of type $2^t$ and size $4t(t-1)$, for $t \in \{4, 5, 6\}$.

CHEE et al.: GROUP DIVISIBLE CODES AND THEIR APPLICATION	3557TABLE II
CODEWORDS OF AN OPTIMAL $(35, 4, [2, 1])_3$-CODE (OF SIZE 291)

| | | | | | | | | |
|---|---|---|---|---|---|---|---|---|
| $\langle 28, 30, 0\rangle$ | $\langle 19, 34, 1\rangle$ | $\langle 5, 8, 6\rangle$ | $\langle 13, 19, 31\rangle$ | $\langle 18, 27, 10\rangle$ | $\langle 1, 18, 2\rangle$ | $\langle 5, 14, 30\rangle$ | $\langle 1, 12, 0\rangle$ | $\langle 24, 28, 11\rangle$ |
| $\langle 3, 28, 27\rangle$ | $\langle 3, 14, 2\rangle$ | $\langle 14, 25, 10\rangle$ | $\langle 4, 32, 0\rangle$ | $\langle 6, 32, 20\rangle$ | $\langle 4, 17, 23\rangle$ | $\langle 7, 8, 11\rangle$ | $\langle 15, 32, 3\rangle$ | $\langle 3, 25, 11\rangle$ |
| $\langle 16, 29, 30\rangle$ | $\langle 24, 30, 9\rangle$ | $\langle 14, 34, 28\rangle$ | $\langle 19, 33, 28\rangle$ | $\langle 27, 31, 1\rangle$ | $\langle 1, 11, 28\rangle$ | $\langle 14, 31, 29\rangle$ | $\langle 3, 4, 6\rangle$ | $\langle 2, 20, 10\rangle$ |
| $\langle 0, 14, 26\rangle$ | $\langle 25, 31, 8\rangle$ | $\langle 24, 27, 26\rangle$ | $\langle 2, 30, 6\rangle$ | $\langle 1, 6, 31\rangle$ | $\langle 28, 29, 20\rangle$ | $\langle 0, 13, 32\rangle$ | $\langle 11, 19, 10\rangle$ | $\langle 24, 34, 17\rangle$ |
| $\langle 14, 15, 1\rangle$ | $\langle 22, 33, 34\rangle$ | $\langle 5, 11, 13\rangle$ | $\langle 8, 30, 26\rangle$ | $\langle 11, 17, 6\rangle$ | $\langle 5, 23, 29\rangle$ | $\langle 10, 34, 11\rangle$ | $\langle 12, 23, 30\rangle$ | $\langle 2, 27, 8\rangle$ |
| $\langle 11, 32, 23\rangle$ | $\langle 5, 34, 20\rangle$ | $\langle 0, 11, 7\rangle$ | $\langle 6, 27, 12\rangle$ | $\langle 9, 32, 19\rangle$ | $\langle 6, 33, 21\rangle$ | $\langle 1, 3, 29\rangle$ | $\langle 6, 26, 5\rangle$ | $\langle 9, 29, 23\rangle$ |
| $\langle 0, 8, 28\rangle$ | $\langle 6, 23, 4\rangle$ | $\langle 10, 28, 5\rangle$ | $\langle 13, 21, 28\rangle$ | $\langle 23, 27, 19\rangle$ | $\langle 23, 24, 13\rangle$ | $\langle 16, 24, 7\rangle$ | $\langle 9, 21, 14\rangle$ | $\langle 0, 29, 1\rangle$ |
| $\langle 10, 29, 19\rangle$ | $\langle 7, 27, 5\rangle$ | $\langle 0, 10, 17\rangle$ | $\langle 12, 18, 8\rangle$ | $\langle 2, 29, 3\rangle$ | $\langle 18, 28, 19\rangle$ | $\langle 24, 25, 18\rangle$ | $\langle 7, 15, 24\rangle$ | $\langle 4, 5, 26\rangle$ |
| $\langle 23, 25, 20\rangle$ | $\langle 11, 18, 34\rangle$ | $\langle 27, 34, 25\rangle$ | $\langle 11, 26, 24\rangle$ | $\langle 4, 9, 24\rangle$ | $\langle 3, 23, 9\rangle$ | $\langle 12, 33, 16\rangle$ | $\langle 19, 22, 23\rangle$ | $\langle 1, 30, 32\rangle$ |
| $\langle 3, 10, 32\rangle$ | $\langle 11, 29, 14\rangle$ | $\langle 2, 15, 25\rangle$ | $\langle 1, 25, 22\rangle$ | $\langle 20, 30, 29\rangle$ | $\langle 20, 21, 31\rangle$ | $\langle 24, 29, 15\rangle$ | $\langle 9, 34, 33\rangle$ | $\langle 0, 5, 21\rangle$ |
| $\langle 1, 20, 34\rangle$ | $\langle 0, 27, 4\rangle$ | $\langle 2, 32, 7\rangle$ | $\langle 18, 26, 15\rangle$ | $\langle 8, 33, 24\rangle$ | $\langle 21, 32, 17\rangle$ | $\langle 19, 21, 29\rangle$ | $\langle 26, 31, 28\rangle$ | $\langle 0, 2, 22\rangle$ |
| $\langle 15, 20, 5\rangle$ | $\langle 1, 33, 23\rangle$ | $\langle 6, 28, 8\rangle$ | $\langle 4, 33, 10\rangle$ | $\langle 16, 34, 2\rangle$ | $\langle 17, 31, 21\rangle$ | $\langle 7, 31, 3\rangle$ | $\langle 14, 18, 22\rangle$ | $\langle 1, 16, 21\rangle$ |
| $\langle 3, 20, 24\rangle$ | $\langle 5, 19, 27\rangle$ | $\langle 10, 15, 22\rangle$ | $\langle 19, 24, 6\rangle$ | $\langle 11, 33, 2\rangle$ | $\langle 14, 17, 7\rangle$ | $\langle 12, 28, 13\rangle$ | $\langle 12, 24, 4\rangle$ | $\langle 16, 23, 18\rangle$ |
| $\langle 15, 23, 11\rangle$ | $\langle 1, 13, 27\rangle$ | $\langle 8, 10, 25\rangle$ | $\langle 1, 24, 8\rangle$ | $\langle 10, 13, 2\rangle$ | $\langle 32, 34, 22\rangle$ | $\langle 22, 31, 9\rangle$ | $\langle 10, 23, 33\rangle$ | $\langle 28, 32, 33\rangle$ |
| $\langle 30, 31, 23\rangle$ | $\langle 26, 33, 0\rangle$ | $\langle 24, 31, 32\rangle$ | $\langle 12, 17, 26\rangle$ | $\langle 14, 27, 21\rangle$ | $\langle 12, 20, 25\rangle$ | $\langle 8, 29, 12\rangle$ | $\langle 8, 17, 2\rangle$ | $\langle 21, 30, 4\rangle$ |
| $\langle 9, 12, 27\rangle$ | $\langle 13, 16, 11\rangle$ | $\langle 26, 29, 27\rangle$ | $\langle 1, 7, 26\rangle$ | $\langle 1, 10, 14\rangle$ | $\langle 7, 18, 9\rangle$ | $\langle 17, 27, 32\rangle$ | $\langle 11, 30, 12\rangle$ | $\langle 18, 33, 25\rangle$ |
| $\langle 21, 24, 22\rangle$ | $\langle 5, 17, 28\rangle$ | $\langle 0, 15, 33\rangle$ | $\langle 17, 33, 9\rangle$ | $\langle 20, 27, 13\rangle$ | $\langle 5, 25, 2\rangle$ | $\langle 10, 24, 20\rangle$ | $\langle 18, 32, 24\rangle$ | $\langle 7, 12, 34\rangle$ |
| $\langle 8, 13, 14\rangle$ | $\langle 7, 28, 14\rangle$ | $\langle 22, 26, 16\rangle$ | $\langle 3, 13, 33\rangle$ | $\langle 13, 15, 12\rangle$ | $\langle 4, 34, 12\rangle$ | $\langle 5, 9, 7\rangle$ | $\langle 17, 30, 10\rangle$ | $\langle 17, 19, 16\rangle$ |
| $\langle 13, 17, 25\rangle$ | $\langle 11, 22, 25\rangle$ | $\langle 2, 23, 17\rangle$ | $\langle 2, 26, 14\rangle$ | $\langle 3, 19, 7\rangle$ | $\langle 2, 19, 4\rangle$ | $\langle 9, 11, 3\rangle$ | $\langle 18, 20, 6\rangle$ | $\langle 26, 32, 1\rangle$ |
| $\langle 12, 22, 14\rangle$ | $\langle 1, 4, 19\rangle$ | $\langle 16, 31, 19\rangle$ | $\langle 4, 7, 2\rangle$ | $\langle 20, 26, 8\rangle$ | $\langle 14, 20, 23\rangle$ | $\langle 9, 15, 18\rangle$ | $\langle 7, 29, 32\rangle$ | $\langle 9, 20, 22\rangle$ |
| $\langle 6, 14, 11\rangle$ | $\langle 7, 10, 16\rangle$ | $\langle 4, 25, 13\rangle$ | $\langle 7, 22, 0\rangle$ | $\langle 10, 26, 4\rangle$ | $\langle 3, 26, 17\rangle$ | $\langle 4, 14, 8\rangle$ | $\langle 4, 11, 27\rangle$ | $\langle 8, 22, 4\rangle$ |
| $\langle 22, 30, 18\rangle$ | $\langle 5, 22, 24\rangle$ | $\langle 14, 19, 9\rangle$ | $\langle 5, 12, 10\rangle$ | $\langle 17, 20, 0\rangle$ | $\langle 14, 16, 12\rangle$ | $\langle 13, 26, 9\rangle$ | $\langle 6, 13, 24\rangle$ | $\langle 16, 27, 3\rangle$ |
| $\langle 6, 16, 9\rangle$ | $\langle 30, 33, 14\rangle$ | $\langle 2, 31, 34\rangle$ | $\langle 25, 26, 12\rangle$ | $\langle 4, 28, 21\rangle$ | $\langle 16, 25, 0\rangle$ | $\langle 13, 18, 23\rangle$ | $\langle 4, 18, 30\rangle$ | $\langle 13, 30, 5\rangle$ |
| $\langle 3, 30, 16\rangle$ | $\langle 19, 26, 30\rangle$ | $\langle 11, 31, 15\rangle$ | $\langle 8, 15, 21\rangle$ | $\langle 0, 19, 20\rangle$ | $\langle 23, 28, 22\rangle$ | $\langle 5, 32, 31\rangle$ | $\langle 9, 28, 26\rangle$ | $\langle 12, 19, 15\rangle$ |
| $\langle 3, 5, 18\rangle$ | $\langle 15, 16, 26\rangle$ | $\langle 3, 12, 31\rangle$ | $\langle 0, 9, 16\rangle$ | $\langle 18, 31, 0\rangle$ | $\langle 29, 34, 31\rangle$ | $\langle 16, 28, 17\rangle$ | $\langle 0, 6, 25\rangle$ | $\langle 8, 19, 18\rangle$ |
| $\langle 6, 7, 30\rangle$ | $\langle 31, 33, 5\rangle$ | $\langle 9, 10, 30\rangle$ | $\langle 1, 9, 17\rangle$ | $\langle 30, 34, 15\rangle$ | $\langle 2, 28, 1\rangle$ | $\langle 8, 32, 27\rangle$ | $\langle 14, 32, 13\rangle$ | $\langle 15, 17, 34\rangle$ |
| $\langle 13, 22, 20\rangle$ | $\langle 8, 9, 31\rangle$ | $\langle 15, 28, 31\rangle$ | $\langle 22, 27, 28\rangle$ | $\langle 5, 16, 33\rangle$ | $\langle 9, 25, 6\rangle$ | $\langle 7, 21, 23\rangle$ | $\langle 8, 23, 1\rangle$ | $\langle 11, 21, 8\rangle$ |
| $\langle 23, 34, 21\rangle$ | $\langle 2, 24, 5\rangle$ | $\langle 15, 27, 30\rangle$ | $\langle 0, 24, 3\rangle$ | $\langle 7, 25, 17\rangle$ | $\langle 21, 25, 15\rangle$ | $\langle 0, 34, 18\rangle$ | $\langle 13, 34, 7\rangle$ | $\langle 3, 22, 15\rangle$ |
| $\langle 12, 21, 6\rangle$ | $\langle 25, 30, 27\rangle$ | $\langle 17, 18, 3\rangle$ | $\langle 6, 15, 19\rangle$ | $\langle 25, 28, 34\rangle$ | $\langle 13, 29, 4\rangle$ | $\langle 1, 5, 15\rangle$ | $\langle 11, 20, 16\rangle$ | $\langle 25, 29, 33\rangle$ |
| $\langle 6, 22, 2\rangle$ | $\langle 10, 21, 27\rangle$ | $\langle 12, 32, 29\rangle$ | $\langle 8, 16, 20\rangle$ | $\langle 4, 15, 29\rangle$ | $\langle 6, 29, 17\rangle$ | $\langle 2, 12, 11\rangle$ | $\langle 4, 31, 20\rangle$ | $\langle 20, 33, 32\rangle$ |
| $\langle 21, 26, 34\rangle$ | $\langle 6, 10, 18\rangle$ | $\langle 3, 8, 34\rangle$ | $\langle 6, 34, 3\rangle$ | $\langle 17, 22, 1\rangle$ | $\langle 2, 9, 13\rangle$ | $\langle 16, 32, 10\rangle$ | $\langle 7, 20, 19\rangle$ | $\langle 18, 29, 5\rangle$ |
| $\langle 10, 31, 12\rangle$ | $\langle 19, 25, 32\rangle$ | $\langle 2, 21, 33\rangle$ | $\langle 0, 23, 31\rangle$ | $\langle 23, 26, 7\rangle$ | $\langle 4, 16, 22\rangle$ | $\langle 22, 29, 21\rangle$ | $\langle 14, 24, 33\rangle$ | $\langle 27, 33, 29\rangle$ |
| $\langle 18, 21, 16\rangle$ | $\langle 7, 33, 13\rangle$ | $\langle 3, 21, 0\rangle$ | | | | | | |

*Proof:* For each $t \in \{4, 5, 6\}$, let $X_t = \mathbb{Z}_{2t}$, and $\mathcal{G}_t = \{\{i, t+i\} : i \in \mathbb{Z}_t\}$. Then $(X_t, \mathcal{G}_t, \mathcal{C}_t)$ is a $[1, 1, 1]$-GDC(3) of type $2^t$ and size $4t(t-1)$, where

- $\mathcal{C}_4$ is the set of all cyclic shifts of the vectors

  $\langle 0, 1, 3\rangle \quad \langle 0, 2, 7\rangle \quad \langle 0, 3, 2\rangle \quad \langle 0, 5, 6\rangle \quad \langle 0, 6, 1\rangle \quad \langle 0, 7, 5\rangle;$

- $\mathcal{C}_5$ is the set of all cyclic shifts of the vectors

  $\langle 0, 1, 2\rangle \quad \langle 0, 2, 6\rangle \quad \langle 0, 3, 9\rangle \quad \langle 0, 4, 3\rangle \quad \langle 0, 6, 8\rangle$
  $\langle 0, 7, 4\rangle \quad \langle 0, 8, 1\rangle \quad \langle 0, 9, 7\rangle;$

- $\mathcal{C}_6$ is the set of all cyclic shifts of the vectors

  $\langle 0, 1, 2\rangle \quad \langle 0, 2, 4\rangle \quad \langle 0, 3, 1\rangle \quad \langle 0, 4, 11\rangle \quad \langle 0, 5, 9\rangle$
  $\langle 0, 7, 10\rangle \quad \langle 0, 8, 5\rangle \quad \langle 0, 9, 8\rangle \quad \langle 0, 10, 3\rangle \quad \langle 0, 11, 7\rangle.$  □

### D. Some $[1, 1, 1]$-GDC(4) and Optimal $(n, 4, [1, 1, 1])_4$-Codes

*Proposition 4.4:* $A_4(n, 4, [1, 1, 1]) = U(n, 4, [1, 1, 1])$ for $n \in \{4, 5, \ldots, 35\} \setminus \{5, 6, 7, 8, 9, 12, 13, 15, 17, 20\}$, and
  i) $A_4(5, 4, [1, 1, 1]) = 6$;
  ii) $A_4(6, 4, [1, 1, 1]) = 11$;
  iii) $A_4(7, 4, [1, 1, 1]) = 16$;
  iv) $A_4(8, 4, [1, 1, 1]) = 23$.

*Proof:* $A_4(n, 4, [1, 1, 1]) = U(n, 4, [1, 1, 1])$ for $n \in \{4, 5, \ldots, 34\} \setminus \{5, 6, 7, 8, 9, 12, 13, 15, 17, 20, 21\}$, has been proven by Chee *et al.* [12]. The remaining values of $A_4(n, 4, [1, 1, 1])$ are established via computer search as follows.

i) The six codewords of an optimal $(5, 4, [1, 1, 1])_4$-code are

$\langle 3, 2, 0\rangle \quad \langle 0, 4, 2\rangle \quad \langle 4, 0, 3\rangle \quad \langle 1, 3, 2\rangle \quad \langle 2, 1, 4\rangle \quad \langle 3, 4, 1\rangle.$

ii) The 11 codewords of an optimal $(6, 4, [1, 1, 1])_4$-code are

$\langle 4, 1, 0\rangle \quad \langle 0, 5, 1\rangle \quad \langle 0, 2, 3\rangle \quad \langle 2, 0, 5\rangle \quad \langle 3, 0, 4\rangle$
$\langle 5, 3, 0\rangle \quad \langle 1, 3, 2\rangle \quad \langle 2, 4, 1\rangle \quad \langle 3, 1, 5\rangle \quad \langle 5, 2, 4\rangle$
$\langle 4, 5, 3\rangle.$

iii) The 16 codewords of an optimal $(7, 4, [1, 1, 1])_4$-code are

$\langle 6, 1, 0\rangle \quad \langle 0, 2, 3\rangle \quad \langle 2, 0, 4\rangle \quad \langle 4, 3, 0\rangle \quad \langle 3, 0, 5\rangle$
$\langle 0, 4, 6\rangle \quad \langle 2, 3, 1\rangle \quad \langle 5, 1, 2\rangle \quad \langle 1, 2, 6\rangle \quad \langle 3, 1, 4\rangle$
$\langle 1, 5, 3\rangle \quad \langle 4, 6, 1\rangle \quad \langle 3, 6, 2\rangle \quad \langle 4, 2, 5\rangle \quad \langle 5, 3, 6\rangle$
$\langle 6, 5, 4\rangle.$

iv) The 23 codewords of an optimal $(8, 4, [1, 1, 1])_4$-code are

$\langle 7, 1, 0\rangle \quad \langle 0, 2, 3\rangle \quad \langle 2, 0, 4\rangle \quad \langle 4, 3, 0\rangle \quad \langle 0, 4, 5\rangle$
$\langle 6, 5, 0\rangle \quad \langle 5, 0, 7\rangle \quad \langle 0, 7, 6\rangle \quad \langle 1, 2, 5\rangle \quad \langle 2, 6, 1\rangle$
$\langle 3, 5, 1\rangle \quad \langle 6, 1, 3\rangle \quad \langle 1, 3, 7\rangle \quad \langle 5, 1, 4\rangle \quad \langle 1, 4, 6\rangle$
$\langle 4, 7, 1\rangle \quad \langle 3, 7, 2\rangle \quad \langle 4, 5, 2\rangle \quad \langle 6, 2, 7\rangle \quad \langle 3, 6, 4\rangle$
$\langle 7, 4, 3\rangle \quad \langle 5, 3, 6\rangle \quad \langle 7, 6, 5\rangle.$

v) The 210 codewords of an optimal $(21, 4, [1, 1, 1])_4$-code are given by the quasi-cyclic shifts with length \3 of

$\langle 2,7,17\rangle$ $\langle 2,9,19\rangle$ $\langle 2,8,6\rangle$ $\langle 3,10,8\rangle$
$\langle 1,8,17\rangle$ $\langle 1,16,2\rangle$ $\langle 3,20,6\rangle$ $\langle 3,11,17\rangle$
$\langle 3,18,14\rangle$ $\langle 3,21,1\rangle$ $\langle 2,11,14\rangle$ $\langle 1,5,21\rangle$
$\langle 2,1,13\rangle$ $\langle 1,18,10\rangle$ $\langle 2,12,20\rangle$ $\langle 3,16,12\rangle$
$\langle 2,21,16\rangle$ $\langle 3,19,2\rangle$ $\langle 1,11,12\rangle$ $\langle 2,5,4\rangle$
$\langle 3,15,9\rangle$ $\langle 1,14,19\rangle$ $\langle 1,3,15\rangle$ $\langle 1,4,6\rangle$
$\langle 2,3,10\rangle$ $\langle 3,4,7\rangle$ $\langle 1,13,7\rangle$ $\langle 2,18,15\rangle$
$\langle 1,20,9\rangle$ $\langle 3,13,5\rangle.$

vi) The 595 codewords of an optimal $(35, 4, [1, 1, 1])_4$-code are given by the cyclic shifts of

$\langle 0,34,1\rangle$ $\langle 0,2,29\rangle$ $\langle 0,3,8\rangle$ $\langle 0,4,23\rangle$
$\langle 0,5,16\rangle$ $\langle 0,6,20\rangle$ $\langle 0,14,7\rangle$ $\langle 0,9,24\rangle$
$\langle 0,11,10\rangle$ $\langle 0,12,21\rangle$ $\langle 0,13,26\rangle$ $\langle 0,15,33\rangle$
$\langle 0,27,17\rangle$ $\langle 0,18,30\rangle$ $\langle 0,19,22\rangle$ $\langle 0,25,31\rangle$
$\langle 0,28,32\rangle.$

*Proposition 4.5:* There exists a $[1, 1, 1]$-GDC(4) of type $4^4$ and size 96.

*Proof:* Let $X = \mathbb{Z}_{16}$ and $\mathcal{G} = \{\{i, 4+i, 8+i, 12+i\} : i \in \mathbb{Z}_4\}$. Then $(X, \mathcal{G}, \mathcal{C})$ is a $[1,1,1]$-GDC(4) of type $4^4$ and size 96, where $\mathcal{C}$ is the set of cyclic shifts of the vectors

$\langle 0,1,2\rangle$ $\langle 0,5,3\rangle$ $\langle 0,6,11\rangle$ $\langle 0,9,15\rangle$ $\langle 0,13,10\rangle$ $\langle 0,14,7\rangle.$ □

*Proposition 4.6:* There exists a $[1, 1, 1]$-GDC(4) of type $6^4$ and size 216.

*Proof:* Let $X = \mathbb{Z}_{24}$ and $\mathcal{G} = \{\{i, 4+i, \ldots, 20+i\} : i \in \mathbb{Z}_4\}$. Then $(X, \mathcal{G}, \mathcal{C})$ is a $[1, 1, 1]$-GDC(4) of type $6^4$ and size 216, where $\mathcal{C}$ is the set of cyclic shifts of the vectors

$\langle 0,19,10\rangle$ $\langle 0,14,13\rangle$ $\langle 0,11,17\rangle$ $\langle 0,6,9\rangle$ $\langle 0,7,21\rangle$
$\langle 0,15,2\rangle$ $\langle 0,3,1\rangle$ $\langle 0,22,5\rangle$ $\langle 0,23,18\rangle$
□

*Proposition 4.7:* There exists a $[1, 1, 1]$-GDC(4) of type $10^4$ and size 600.

*Proof:* Let $X = \mathbb{Z}_{40}$ and $\mathcal{G} = \{\{i, 4+i, \ldots, 36+i\} : i \in \mathbb{Z}_4\}$. Then $(X, \mathcal{G}, \mathcal{C})$ is a $[1,1,1]$-GDC(4) of type $10^4$ and size 600, where $\mathcal{C}$ is the set of cyclic shifts of the vectors

$\langle 0,26,5\rangle$ $\langle 0,2,29\rangle$ $\langle 0,7,38\rangle$ $\langle 0,18,25\rangle$ $\langle 0,39,10\rangle$
$\langle 0,27,17\rangle$ $\langle 0,34,9\rangle$ $\langle 0,35,37\rangle$ $\langle 0,3,21\rangle$ $\langle 0,15,1\rangle$
$\langle 0,19,13\rangle$ $\langle 0,23,22\rangle$ $\langle 0,30,33\rangle$ $\langle 0,11,6\rangle$ $\langle 0,31,14\rangle.$
□

*Proposition 4.8:* There exists a $[1, 1, 1]$-GDC(4) of type $1^{20}11^1$ and size 410.

*Proof:* Adjoin one point to a $[1, 1, 1]$-GDC(4) of type $10^3$ and size 300 (which exists by Theorem 7.1) using an optimal $(11, 4, [1, 1, 1])_4$-code as ingredient to obtain a $(31, 4, [1, 1, 1])_4$-code of size 465. This code contains an optimal $(11, 4, [1, 1, 1])_4$-code. Removing this code gives a $[1, 1, 1]$-GDC(4) of type $1^{20}11^1$ of size 410.

## V. DETERMINING THE VALUE OF $A_3(N, 4, [2, 1])$

The value of $A_3(n, 4, [2, 1])$ has been completely determined for the cases $n \equiv 0 \pmod 2$ [19] and $n \equiv 1 \pmod 4$ [12]. In this section, we solve the case $n \equiv 3 \pmod 4$ completely. We consider three congruence classes of $n$.

*Proposition 5.1:* $A_3(n, 4, [2, 1]) = U(n, 4, [2, 1])$ for all $n \equiv 3 \pmod{12}, n \geq 39$.

*Proof:* Let $t \geq 3$. Inflate a $\{3\}$-GDD of type $6^t$ (which exists by Theorem 3.9) by two using a $[2, 1]$-GDC(4) of type $2^3$ and size six (which exists by Example 3.1) as ingredient. This gives a $[2, 1]$-GDC(4) of type $12^t$. Adjoining three points to this $[2, 1]$-GDC(4) of type $12^t$ using an optimal $(15, 4, [2, 1])_3$-code of size 50 (which exists by Proposition 4.2) and a $[2, 1]$-GDC(4) of type $1^{12}3^1$ and size 49 (which exists by Example 3.4), gives a $(12t + 3, 4, [2, 1])_3$-code of size

$$\frac{6}{\binom{3}{2}}\left(\binom{6t}{2} - t\binom{6}{2}\right) + 50 + 49(t-1)$$
$$= 36t^2 + 13t + 1$$
$$= U(12t + 3, 4, [2, 1])$$

which is optimal.

*Proposition 5.2:* $A_3(n, 4, [2, 1]) = U(n, 4, [2, 1])$ for all $n \equiv 7 \pmod{12}, n \geq 43$.

*Proof:* Let $t \geq 3$. Take a $\{3\}$-GDD of type $6^t 2^1$ (which exists by Theorem 3.9) as the master GDD and apply the Fundamental Construction with $\omega(\,\cdot\,) = 2$. This gives a $[2, 1]$-GDC(4) of type $12^t 4^1$. The required ingredient $[2, 1]$-GDC(4) of type $2^3$ and size six exists by Example 3.1. Adjoining three points to this $[2, 1]$-GDC(4) of type $12^t 4^1$ using an optimal $(7, 4, [2, 1])_3$-code of size nine (which exists by Proposition 4.2) and a $[2, 1]$-GDC(4) of type $1^{12}3^1$ and size 49(which exists by Example 3.4), gives a $(12t + 7, 4, [2, 1])_3$-code of size

$$\frac{6}{\binom{3}{2}}\left(\binom{6t+2}{2} - t\binom{6}{2} - \binom{2}{2}\right) + 49t + 9$$
$$= 36t^2 + 37t + 9$$
$$= U(12t + 7, 4, [2, 1]).$$

which is optimal.

*Proposition 5.3:* $A_3(n, 4, [2, 1]) = U(n, 4, [2, 1])$ for all $n \equiv 11 \pmod{12}, n \geq 47$.

*Proof:* Let $t \geq 3$. Take a $\{3\}$-GDD of type $6^t 4^1$ (which exists by Theorem 3.9) as the master GDD and apply the Fundamental Construction with $\omega(\,\cdot\,) = 2$. This gives a $[2, 1]$-GDC(4) of type $12^t 8^1$. The required ingredient $[2, 1]$-GDC(4) of type $2^3$ and size six exists by Example 3.1. Adjoining three points to this $[2, 1]$-GDC(4) of type $12^t 8^1$ using an optimal $(11, 4, [2, 1])_3$-code of size 25 (which exists by Proposition 4.2) and a $[2, 1]$-GDC(4) of type $1^{12}3^1$ and size 49 (which exists by Example 3.4), gives a $(12t + 11, 4, [2, 1])_3$-code of size

$$\frac{6}{\binom{3}{2}}\left(\binom{6t+4}{2} - t\binom{6}{2} - \binom{4}{2}\right) + 49t + 25$$
$$= 36t^2 + 61t + 25$$
$$= U(12t + 11, 4, [2, 1])$$

which is optimal.

TABLE III
SOME {4, 5, 6}-GDDs

| Type | $5^4 2^1$ | $5^4 3^1$ | $5^5$ | $5^5 2^1$ | $5^5 4^1$ | $7^4 3^1$ | $16^4 15^1$ | $17^4 15^1$ | $18^4 14^1$ |
|---|---|---|---|---|---|---|---|---|---|
| Number of blocks of size 4 | 15 | 10 | 0 | 0 | 0 | 28 | 16 | 34 | 72 |
| Number of blocks of size 5 | 10 | 15 | 25 | 15 | 5 | 21 | 240 | 255 | 252 |
| Number of blocks of size 6 | 0 | 0 | 0 | 10 | 20 | 0 | 0 | 0 | 0 |

TABLE IV
SOME [1, 1, 1]-GDC(3)

| Type | $10^4 4^1$ | $10^4 6^1$ | $10^5$ | $10^5 4^1$ | $10^5 8^1$ | $14^4 6^1$ | $32^4 30^1$ | $34^4 30^1$ | $36^4 28^1$ |
|---|---|---|---|---|---|---|---|---|---|
| Length | 44 | 46 | 50 | 54 | 58 | 62 | 158 | 166 | 172 |
| Size | 1520 | 1680 | 2000 | 2400 | 2800 | 3024 | 19968 | 22032 | 23616 |
| $y$ | 0 | 1 | 1 | 0 | 1 | 0 | 0 | 1 | 1 |
| Size of $(n, 3, [1, 1, 1])_4$-code after adjoining $y$ points | 1892 | 2162 | 2550 | 2862 | 3422 | 3782 | 24806 | 27722 | 29756 |

*Corollary 5.1:* $A_3(n, 4, [2, 1]) = U(n, 4, [2, 1])$ for all $n \equiv 3 \pmod{4}$.

*Proof:* Follows from Propositions 4.2, 5.1, 5.2, and 5.3. □

## VI. DETERMINING THE VALUE OF $A_4(N, 3, [1, 1, 1])$

*Proposition 6.1:* There exist $\{4, 5, 6\}$-GDD's of the types listed in Table III.

*Proof:* For types $5^4 2^1, 5^4 3^1$, and $5^5$, truncate points from a TD(5, 5). For types $5^5 2^1$ and $5^5 4^1$, truncate points from a TD(6, 5). For type $7^4 3^1$, truncate points from a TD(5, 7). For type $16^4 15^1$, truncate points from a TD(5, 16). For type $17^4 15^1$, truncate points from a TD(5, 17). For type $18^4 14^1$, truncate points from a TD(5, 18).

*Proposition 6.2:* $A_4(n, 3, [1, 1, 1]) = U(n, 3, [1, 1, 1])$ for $n \in \{44, 47, 51, 54, 59, 62, 158, 167, 173\}$.

*Proof:* Inflate a $\{4, 5, 6\}$-GDD of type $g_1^t g_2^1$ (provided by Proposition 6.1) by two to obtain a $[1, 1, 1]$-GDC(3) of type $(2g_1)^t (2g_2)^1$. The required ingredient $[1, 1, 1]$-GDC(3) of types $2^4, 2^5, 2^6$ having size $48, 80, 120$, respectively, all exist by Proposition 4.3. The size of the resulting $[1, 1, 1]$-GDC(3) is given in Table IV. Now adjoin $y$ points to this $[1, 1, 1]$-GDC(3), where $y$ is given in Table IV. The $(n, 3, [1, 1, 1])_4$-codes so obtained are optimal.

## VII. DETERMINING THE VALUE OF $A_4(N, 4, [1, 1, 1])$

Chee et al. [12, Lemma 21] have proven that if $n$ is odd and $A_4(n, 4, [1, 1, 1]) = U(n, 4, [1, 1, 1])$, then $A_4(n-1, 4, [1, 1, 1]) = U(n-1, 4, [1, 1, 1])$. Hence, we focus first on establishing results for $(n, 4, [1, 1, 1])_4$-codes of odd lengths. We begin with some general constructions for GDCs and optimal codes.

### A. General Constructions

*Theorem 7.1:* There exists a $[1, 1, 1]$-GDC(4) of type $g^3$ and size $3g^2$, for all $g \geq 3$.

*Proof:* Let $L$ be a Latin square of side $g$ with entries from the set $S = \{0, \ldots, g - 1\}$, and with its rows and columns also indexed by $S$. Define the bijection $\pi_i : s \mapsto s + \pmod{g}$, for $s \in S$, and let $L_i$ denote the Latin square obtained from $L$ by replacing each entry $s$ of $L$ by $\pi_i(s)$. It is easy to see that $L_0, L_1$, and $L_2$ are *pairwise disjoint*, that is, for each $(r, c) \in S \times S$, the entries in cell $(r, c)$ of $L_0, L_1$, and $L_2$ are all distinct.

Define the $[1, 1, 1]$-GDC $(X, \mathcal{G}, \mathcal{C}_0 \cup \mathcal{C}_1 \cup \mathcal{C}_2)$ of type $g^3$, where

$X = \{0, \ldots, 3g - 1\}$
$\mathcal{G} = \{\{0, \ldots, g - 1\}, \{g, \ldots, 2g - 1\}, \{2g, \ldots, 3g - 1\}\}$
$\mathcal{C}_0 = \{\langle r, c + g, s + 2g \rangle :$
  $0 \leq r, c < g$ and $s$ is the entry in cell $(r, c)$ of $L_0\}$
$\mathcal{C}_1 = \{\langle s + 2g, r, c + g \rangle :$
  $0 \leq r, c < g$ and $s$ is the entry in cell $(r, c)$ of $L_1\}$
$\mathcal{C}_2 = \{\langle c + g, s + 2g, r \rangle :$
  $0 \leq r, c < g$ and $s$ is the entry in cell $(r, c)$ of $L_2\}$.

We claim that $\mathcal{C}_0 \cup \mathcal{C}_1 \cup \mathcal{C}_2$ is a code of distance four. Indeed, for distinct $\mathsf{u}, \mathsf{v} \in \mathcal{C}_i, i \in \{0, 1, 2\}$, the property of a Latin square ensures that $d_H(\mathsf{u}, \mathsf{v}) \geq 4$. If $\mathsf{u} \in \mathcal{C}_i$ and $\mathsf{v} \in \mathcal{C}_j, 0 \leq i < j \leq 2$, then since $L_i$ and $L_j$ are disjoint, then $\mathsf{u}$ and $\mathsf{v}$ can share at most two nonzero coordinates. However, these coordinates must receive different values by our construction. Thus, $(X, \mathcal{G}, \mathcal{C}_0 \cup \mathcal{C}_1 \cup \mathcal{C}_2)$ is a $[1, 1, 1]$-GDC(4) of type $g^3$ and size $3g^2$.

*Corollary 7.1 (Tripling Construction):* Let $n$ be an odd positive integer. If $A_4(n, 4, [1, 1, 1]) = U(n, 4, [1, 1, 1])$, then we have

$$A_4(3n, 4, [1, 1, 1]) = U(3n, 4, [1, 1, 1]),$$

and

$$A_4(3(n-1) + 1, 4, [1, 1, 1]) = U(3(n-1) + 1, 4, [1, 1, 1]).$$

*Proof:* Fill in the groups of a $[1, 1, 1]$-GDC(4) of type $n^3$ and size $3n^2$, which exists by Theorem 7.1, to obtain a $(3n, 4, [1, 1, 1])_4$-code of size

$$3n^2 + 3U(n, 4, [1, 1, 1])$$
$$= 3n^2 + 3\binom{n}{2}$$
$$= \binom{3n}{2}$$
$$= U(3n, 4, [1, 1, 1]).$$

Hence, we have $A_4(3n, 4, [1, 1, 1]) = U(3n, 4, [1, 1, 1])$. Similarly, to prove $A_4(3(n-1) + 1, 4, [1, 1, 1]) = U(3(n-1) + 1, 4, [1, 1, 1])$, we adjoin a point and fill in the groups of a $[1, 1, 1]$-GDC(4) of type $(n-1)^3$ and size $3(n-1)^2$ with an $(n, 4, [1, 1, 1])_4$-code of size $U(n, 4, [1, 1, 1])$. □

*Theorem 7.2 (Prime Power Construction):* Let $n \equiv 3 \pmod 4$ be a prime power with $n \geq 11$. Suppose there is a generator $\alpha$ in the finite field $\mathbb{F}_n$ such that the following conditions hold:
1) $\alpha - 1$ is a quadratic residue;
2) $\alpha^2 - \alpha + 1 \neq 0$.

Then $A_4(n, 4, [1, 1, 1]) = U(n, 4, [1, 1, 1])$. In particular, we have $A_4(n, 4, [1, 1, 1]) = U(n, 4, [1, 1, 1])$ for $n \in \{43, 47, 59, 67, 71, 83, 107\}$.

*Proof:* Let $\alpha$ be the generator of the finite field $\mathbb{F}_n$ satisfying the above two conditions. Let $C = \langle 0, 1, \alpha \rangle$ and consider the code $\mathcal{C} = \{\alpha^{2i} C + x : 0 \leq i < (n-1)/2 \text{ and } x \in \mathbb{F}_n\}$. Obviously, $\mathcal{C}$ is a code of length $n$ and constant composition $[1, 1, 1]$. We show that $\mathcal{C}$ has distance four and size $n(n-1)/2$. Since $\alpha$ meets the above two conditions, it can be easily checked that any two codewords can share at most two nonzero coordinates. Suppose $\mathsf{u} = \alpha^{2i} C + x$ and $\mathsf{v} = \alpha^{2j} C + y$, where $(i, x) \neq (j, y)$

$$\mathsf{u} = \langle x, \alpha^{2i} + x, \alpha^{2i+1} + x \rangle$$
$$\mathsf{v} = \langle y, \alpha^{2j} + y, \alpha^{2j+1} + y \rangle.$$

If $x = y$, then $\alpha^{2i} + x \neq \alpha^{2j} + y$ and $\alpha^{2i+1} + x \neq \alpha^{2j+1} + y$, unless $i = j$, a contradiction. Hence, if $x = y$, then $d_H(\mathsf{u}, \mathsf{v}) = 4$.

If $x \neq y$, then if

$$\alpha^{2i} + x = \alpha^{2j} + y \quad (1)$$

and

$$\alpha^{2i+1} + x = \alpha^{2j+1} + y \quad (2)$$

we have $\alpha^{2i} - \alpha^{2i+1} = \alpha^{2j} - \alpha^{2j-1}$, which gives $\alpha^{2i}(1 - \alpha) = \alpha^{2j}(1 - \alpha)$, implying $i = j$. However, $i = j$ implies that (1) and (2) can hold only if $x = y$, a contradiction. Therefore, at most one of (1) and (2) can hold. Consequently, if $x \neq y$, then $d_H(\mathsf{u}, \mathsf{v}) \geq 4$.

The proof above also shows that all codewords in $\mathcal{C}$ are distinct, for otherwise there would be two codewords that are distance zero apart. It follows that the size of $\mathcal{C}$ is $n(n-1)/2$. Finally, for primes $n = 43, 47, 59, 67, 71, 83, 107$, we can take $\alpha = 26, 5, 2, 2, 7, 2, 2$, respectively. □

*Example 7.1:* The 1081 codewords of an (optimal) $(47, 4, [1, 1, 1])_4$-code are given by $5^{2i} C + x, 0 \leq i < 23, 0 \leq x < 47$, where $C = \langle 0, 1, 5 \rangle$.

The idea of Theorem 7.2 can be extended to a computational search procedure for optimal $(n, 4, [1, 1, 1])_4$-codes when $n \equiv 1 \pmod 4$. We developed an algorithm to look for $C_1 = \langle 0, 1, \alpha \rangle, C_2, \ldots, C_t$ such that

$$\mathcal{C} = \{\alpha^{2ti} C_j + x : 0 \leq i \leq (n-1)/2t, x \in \mathbb{F}_n, \text{ and } j \in [t]\}$$

is an optimal $(n, 4, [1, 1, 1])_4$-code. We call $C_1, \ldots, C_t$ the *base codewords*. We have been successful in determining optimal $(n, 4, [1, 1, 1])_4$-codes in the following instances.

*Proposition 7.1:* $A_4(n, 4, [1, 1, 1]) = U(n, 4, [1, 1, 1])$ for $n \in \{37, 41, 53, 61, 89\}$.

*Proof:*
i) For $n = 37$, take as base blocks $C_1 = \langle 0, 1, 2 \rangle$ and $C_2 = \langle 0, 5, 3 \rangle$.
ii) For $n = 41$, take as base blocks $C_1 = \langle 0, 1, 2 \rangle, C_2 = \langle 0, 3, 11 \rangle, C_3 = \langle 0, 5, 31 \rangle$, and $C_4 = \langle 0, 6, 35 \rangle$.
iii) For $n = 53$, take as base blocks $C_1 = \langle 0, 1, 2 \rangle$ and $C_2 = \langle 0, 5, 17 \rangle$.
iv) For $n = 61$, take as base blocks $C_1 = \langle 0, 1, 2 \rangle$ and $C_2 = \langle 0, 6, 4 \rangle$.
v) For $n = 89$, take as base blocks $C_1 = \langle 0, 1, 3 \rangle, C_2 = \langle 0, 5, 15 \rangle, C_3 = \langle 0, 19, 57 \rangle$, and $C_4 = \langle 0, 33, 9 \rangle$. □

*B. Odd Lengths: $n \equiv 1$ (Mod 6)*

We first consider the easy case of $n \equiv 1 \pmod 6$.

*Proposition 7.2:* $A_4(n, 4, [1, 1, 1]) = U(n, 4, [1, 1, 1])$ for all $n \equiv 1 \pmod 6, n \geq 19$, and $n \neq 49$.

*Proof:* For $n \equiv 1 \pmod{18}, n \geq 55$, inflate a $\{3\}$-GDD of type $6^{(n-1)/18}$ (which exists by Theorem 3.9) by three to obtain a $[1, 1, 1]$-GDC(4) of type $18^{(n-1)/18}$. The required ingredient $[1, 1, 1]$-GDC(4) of type $3^3$ and size 27 exists by Theorem 7.1. Adjoining one point to this $[1, 1, 1]$-GDC(4) of type $18^{(n-1)/18}$ using an optimal $(19, 4, [1, 1, 1])_4$-code of size 171 (which exists by Proposition 4.4) gives an $(n, 4, [1, 1, 1])_4$-code of size

$$\frac{27}{\binom{3}{2}} \left( \binom{(n-1)/3}{2} - \frac{n-1}{18} \binom{6}{2} \right) + 171 \cdot \frac{n-1}{18}$$
$$= \frac{n(n-1)}{2}$$
$$= U(n, 4, [1, 1, 1]).$$

For $n \equiv 7 \pmod{18}, n \geq 79$, inflate a $\{3\}$-GDD of type $6^{(n-25)/18} 8^1$ (which exists by Theorem 3.9) by three to obtain a $[1, 1, 1]$-GDC(4) of type $18^{(n-25)/18} 24^1$. The required ingredient $[1, 1, 1]$-GDC(4) of type $3^3$ and size 27 exists by Theorem 7.1. Adjoining one point to this $[1, 1, 1]$-GDC(4) of type $18^{(n-25)/18} 24^1$ using an optimal $(19, 4, [1, 1, 1])_4$-code of size 171 and an optimal $(25, 4, [1, 1, 1])_4$-code of size 300 (which exist by Proposition 4.4) gives an $(n, 4, [1, 1, 1])_4$-code of size

$$\frac{27}{\binom{3}{2}} \left( \binom{(n-1)/3}{2} - \frac{n-25}{18} \binom{6}{2} \right.$$
$$\left. - \binom{8}{2} \right) + 171 \cdot \frac{n-25}{18} + 300$$
$$= \frac{n(n-1)}{2}$$
$$= U(n, 4, [1, 1, 1]).$$

For $n \equiv 13 \pmod{18}, n \geq 85$, inflate a $\{3\}$-GDD of type $6^{(n-31)/18} 10^1$ (which exists by Theorem 3.9) by three to obtain a $[1, 1, 1]$-GDC(4) of type $18^{(n-31)/18} 30^1$. The required ingredient $[1, 1, 1]$-GDC(4) of type $3^3$ and size 27 exists by Theorem 7.1. Adjoining one point to this $[1, 1, 1]$-GDC(4) of type $18^{(n-31)/18} 30^1$ using an optimal $(19, 4, [1, 1, 1])_4$-code of size 171 and an optimal $(31, 4, [1, 1, 1])_4$-code of size 465 (which exist by Proposition 4.4) gives an $(n, 4, [1, 1, 1])_4$-code of size

$$\frac{27}{\binom{3}{2}}\left(\binom{(n-1)/3}{2} - \frac{n-31}{18}\binom{6}{2} - \binom{10}{2}\right) + 171 \cdot \frac{n-31}{18} + 465$$
$$= \frac{n(n-1)}{2}$$
$$= U(n, 4, [1, 1, 1]).$$

The above establishes $A_4(n, 4, [1, 1, 1]) = U(n, 4, [1, 1, 1])$ for all $n \equiv 1 \pmod{6}$, $n \geq 19$, $n \notin \{19, 25, 31, 37, 43, 49, 61, 67\}$. For $n \in \{19, 25, 31\}$, we have $A_4(n, 4, [1, 1, 1]) = U(n, 4, [1, 1, 1])$ by Proposition 4.4. For $n \in \{37, 61\}$, we have $A_4(n, 4, [1, 1, 1]) = U(n, 4, [1, 1, 1])$ by Proposition 7.1. For $n \in \{43, 67\}$, we have $A_4(n, 4, [1, 1, 1]) = U(n, 4, [1, 1, 1])$ by Theorem 7.2. □

*Corollary 7.2:* There exists a $[1, 1, 1]$-GDC(4) of type $1^t 19^1$ and size $t(t + 37)/2$, for $t \in \{36, 60\}$.

*Proof:* Observe that the codes of length $n$ constructed in Proposition 7.2 contains a $(19, 4, [1, 1, 1])_4$-code of size 171. Removing this $(19, 4, [1, 1, 1])_4$-code of size 171 gives a $[1, 1, 1]$-GDC(4) of type $1^{n-19} 19^1$. Taking $n \in \{55, 79\}$ then gives the required result. □

### C. Odd Lengths: $n \equiv 1 \pmod{4}$

We establish a general construction for optimal $(n, 4, [1, 1, 1])_4$-codes, $n \equiv 1 \pmod{4}$, from $\{3, 4\}$-GDDs.

*Theorem 7.3:* If there exists a $\{3, 4\}$-GDD of type $g_1^{s_1} \cdots g_t^{s_t}$, and there exists a $(4g_i + 1, 4, [1, 1, 1])_4$-code of size $U(4g_i + 1, 4, [1, 1, 1])$ for each $i \in \{1, \ldots, t\}$, then there exists a $(4\sum_{i=1}^t s_i g_i + 1, 4, [1, 1, 1])_4$-code of size $U(4\sum_{i=1}^t s_i g_i + 1, 4, [1, 1, 1])$.

*Proof:* Let $(X, \mathcal{G}, \mathcal{A})$ be a $\{3, 4\}$-GDD of type $g_1^{s_1} \cdots g_t^{s_t}$ with $x$ blocks of size three and $y$ blocks of size four. Then $x$ and $y$ satisfy

$$3x + 6y = \binom{\sum_{i=1}^t s_i g_i}{2} - \sum_{i=1}^t s_i \binom{g_i}{2}.$$

Now inflate $(X, \mathcal{G}, \mathcal{A})$ by four using a $[1, 1, 1]$-GDC(4) of type $4^3$ and size 48 (which exists by Theorem 7.1), and a $[1, 1, 1]$-GDC(4) of type $4^4$ and size 96 (which exists by Proposition 4.5) as ingredients. This gives a $[1, 1, 1]$-GDC(4) of type $(4g_1)^{s_1} \cdots (4g_t)^{s_t}$ having size $48x + 96y$. Now adjoin one point to this GDC to obtain a $(4\sum_{i=1}^t s_i g_i + 1, 4, [1, 1, 1])_4$-code of size

$$48x + 96y + \sum_{i=1}^t s_i (4g_i + 1)(4g_i)/2$$
$$= 16\binom{\sum_{i=1}^t s_i g_i}{2}$$
$$- 16 \sum_{i=1}^t s_i \binom{g_i}{2} + \sum_{i=1}^t s_i \binom{4g_i + 1}{2}$$
$$= 16\binom{\sum_{i=1}^t s_i g_i}{2} + 10 \sum_{i=1}^t s_i g_i$$
$$= \binom{4\sum_{i=1}^t s_i g_i + 1}{2}$$
$$= U\left(4\sum_{i=1}^t s_i g_i + 1, 4, [1, 1, 1]\right). \quad \square$$

To apply Theorem 7.3, we require classes of $\{3, 4\}$-GDDs provided below.

*Proposition 7.3:* There exists a $\{3, 4\}$-GDD of the following types:
  i) $7^3 8^1, 9^3 8^1, 13^3 8^1, 15^3 8^1$;
  ii) $8^4 9^1, 7^6 8^1 9^1$;
  iii) $6^t u^1$, for $u \in \{0, 7, 8, 9, 10\}$ and $t \geq 3$;
  iv) $6^t 7^3 8^1$, for $t \geq 7$.

*Proof:*
  i) Take a TD$(4, 8)$ and pick a block $\{a, b, c, d\}$. Removing this block and deleting each of the points $a, b, c$ from the groups and blocks where they occur gives a $\{3, 4\}$-GDD of type $7^3 8^1$. $\{3, 4\}$-GDDs of types $9^3 8^1, 13^3 8^1, 15^3 8^1$ can be obtained by truncating a group of a TD$(4, g)$.
  ii) The existence of $\{3, 4\}$-GDD's of types $8^4 9^1$ and $7^6 8^1 9^1$ has already being established in Proposition 4.1.
  iii) By Theorem 3.9, a $\{3\}$-GDD of type $6^t u^1$ exists for all $u \in \{0, 8, 10\}$ and $t \geq 3$. By Theorem 3.12, a $\{4\}$-GDD of type $6^t 9^1$ exists for all $t \geq 4$. Truncate two points from the group of size nine of this GDD to obtain a $\{3, 4\}$-GDD of type $6^t 7^1$ for $t \geq 4$. What remains is to show the existence of $\{3, 4\}$-GDDs of types $6^3 7^1$ and $6^3 9^1$. To construct these, let $(X, \mathcal{G}, \mathcal{A})$ be a Kirkman frame of type $6^4$, which exists by Theorem 3.2, and let $\mathcal{A}' \subseteq \mathcal{A}$ be a holey parallel class with $\cup_{A \in \mathcal{A}'} A = X \setminus G$, for some $G \in \mathcal{G}$. By Theorem 3.3, adding one point to this frame gives a $\{3, 4\}$-GDD of type $6^3 7^1$, while adding three points gives a $\{3, 4\}$-GDD of type $6^3 9^1$.
  iv) First, we prove the existence of a $\{3, 4\}$-GDD of type $6^t 29^1$. For odd $t \geq 7$, take a $\{4\}$-GDD of type $3^t(3(t - 1)/2)^1$ (which exists by Theorem 3.1) as the master GDD and apply the Fundamental Construction with weight function assigning weight two to the points in the groups of size three, and weight in $\{0, 2, 3, 4\}$ to the points in the remaining groups to obtain a $\{3, 4\}$-GDD of type $6^t 29^1$. The required ingredient $\{3, 4\}$-GDD's of types $2^3 u^1$ for $u \in \{0, 2, 4\}$ exist by Theorem 3.9 and an ingredient $\{3, 4\}$-GDD of type $2^3 3^1$ exists by taking a TD$(4, 3)$ and truncating a block by three points. For even $t = 2k \geq 8$, take a $\{3\}$-RGDD $(X, \mathcal{G}, A)$ of type $2^{3k}$ (which exists by Theorem 3.1), having $\mathcal{A} = \mathcal{A}_1 \cup \cdots \cup \mathcal{A}_{3k-1}$ as its partition into parallel classes. Now let $\infty_1, \ldots, \infty_{3k-1} \notin X$ and consider

$$X' = X \cup \{\infty_1, \ldots, \infty_{3k-1}\}$$
$$\mathcal{G}' = \mathcal{A}_{3k-1} \cup \{\{\infty_1, \ldots, \infty_{3k-1}\}\}$$
$$\mathcal{A}' = \left(\cup_{i=1}^{3k-2}\{A \cup \{\infty_i\} : A \in \mathcal{A}_i\}\right)$$
$$\cup \{G \cup \{\infty_{3k-1}\} : G \in \mathcal{G}\}.$$

Then $(X', \mathcal{G}', \mathcal{A}')$ is a $\{3, 4\}$-GDD of type $3^{2k}(3k - 1)^1$.

As for the case when $t$ is odd above, apply the Fundamental Construction to obtain a $\{3,4\}$-GDD of type $6^t 29^1$.

Finally, fill in the group of size 29 of a $\{3,4\}$-GDD of type $6^t 29^1$ by a $\{3,4\}$-GDD of type $7^3 8^1$ (from i) above) to obtain a $\{3,4\}$-GDD of type $6^t 7^3 8^1$.

*Theorem 7.4:* $A_4(n,4,[1,1,1]) = U(n,4,[1,1,1])$ for all $n \equiv 1 \pmod{4}, n \geq 37$.

*Proof:* Propositions 4.4 (with Corollary 7.1), 7.1, 7.2, and 7.3 (with Theorem 7.3) establish the theorem for all $n \equiv 1 \pmod{4}, n \geq 37$, and $n \notin \{45, 49, 65, 77\}$.

For $n = 77$, inflate a $\{3\}$-GDD of type $1^7$ (which exists by Theorem 3.9) by 11 using an ingredient $[1,1,1]$-GDC(4) of type $11^3$ and size 363 (which exists by Theorem 7.1) to obtain a $[1,1,1]$-GDC(4) of type $11^7$ and size 2541. Now fill in the groups of this GDC with optimal $(11,4,[1,1,1])_4$-codes to obtain a $(77,4,[1,1,1])_4$-code of size 2926, which is optimal.

The remaining values of $A_4(n,4,[1,1,1])$ with $n \in \{45, 49, 65\}$ are established via computer search.

i) The 990 codewords of an optimal $(45,4,[1,1,1])_4$-code are given by the cyclic shifts of

$\langle 0,2,35 \rangle$ $\langle 0,22,24 \rangle$ $\langle 0,16,30 \rangle$ $\langle 0,32,40 \rangle$
$\langle 0,1,6 \rangle$ $\langle 0,11,12 \rangle$ $\langle 0,14,34 \rangle$ $\langle 0,4,31 \rangle$
$\langle 0,42,7 \rangle$ $\langle 0,19,41 \rangle$ $\langle 0,20,29 \rangle$ $\langle 0,27,44 \rangle$
$\langle 0,17,28 \rangle$ $\langle 0,21,18 \rangle$ $\langle 0,8,23 \rangle$ $\langle 0,15,36 \rangle$
$\langle 0,33,37 \rangle$ $\langle 0,9,3 \rangle$ $\langle 0,10,26 \rangle$ $\langle 0,5,43 \rangle$
$\langle 0,38,25 \rangle$ $\langle 0,39,13 \rangle$.

ii) The 1176 codewords of an optimal $(49,4,[1,1,1])_4$-code are given by the cyclic shifts of

$\langle 0,48,11 \rangle$ $\langle 0,2,5 \rangle$ $\langle 0,3,31 \rangle$ $\langle 0,23,47 \rangle$
$\langle 0,14,37 \rangle$ $\langle 0,22,6 \rangle$ $\langle 0,44,39 \rangle$ $\langle 0,29,46 \rangle$
$\langle 0,4,40 \rangle$ $\langle 0,38,8 \rangle$ $\langle 0,17,21 \rangle$ $\langle 0,9,1 \rangle$
$\langle 0,41,35 \rangle$ $\langle 0,24,42 \rangle$ $\langle 0,12,34 \rangle$ $\langle 0,7,16 \rangle$
$\langle 0,33,13 \rangle$ $\langle 0,19,26 \rangle$ $\langle 0,10,20 \rangle$ $\langle 0,28,27 \rangle$
$\langle 0,43,45 \rangle$ $\langle 0,15,30 \rangle$ $\langle 0,18,32 \rangle$ $\langle 0,36,25 \rangle$.

iii) The 2080 codewords of an optimal $(65,4,[1,1,1])_4$-code are given by the cyclic shifts of

$\langle 0,54,33 \rangle$ $\langle 0,63,40 \rangle$ $\langle 0,47,14 \rangle$ $\langle 0,32,35 \rangle$
$\langle 0,42,28 \rangle$ $\langle 0,5,60 \rangle$ $\langle 0,37,45 \rangle$ $\langle 0,56,61 \rangle$
$\langle 0,20,50 \rangle$ $\langle 0,15,16 \rangle$ $\langle 0,8,27 \rangle$ $\langle 0,22,11 \rangle$
$\langle 0,55,39 \rangle$ $\langle 0,6,12 \rangle$ $\langle 0,52,23 \rangle$ $\langle 0,44,48 \rangle$
$\langle 0,49,31 \rangle$ $\langle 0,3,18 \rangle$ $\langle 0,34,21 \rangle$ $\langle 0,53,13 \rangle$
$\langle 0,19,10 \rangle$ $\langle 0,30,2 \rangle$ $\langle 0,36,9 \rangle$ $\langle 0,38,62 \rangle$
$\langle 0,4,57 \rangle$ $\langle 0,51,58 \rangle$ $\langle 0,25,59 \rangle$ $\langle 0,1,64 \rangle$
$\langle 0,7,29 \rangle$ $\langle 0,17,43 \rangle$ $\langle 0,24,41 \rangle$ $\langle 0,26,46 \rangle$.

### D. Odd Lengths: $n \equiv 3 \pmod{4}$

First, we settle the case $n \equiv 3 \pmod{12}$.

*Proposition 7.4:* $A_4(n,4,[1,1,1]) = U(n,4,[1,1,1])$ for all $n \equiv 3 \pmod{12}, n \geq 39$.

*Proof:* Apply Corollary 7.1 with Proposition 4.4 and Theorem 7.4 to establish the theorem for $n \equiv 3 \pmod{12}, n \geq 39$ and $n \notin \{39, 51\}$.

The remaining values of $A_4(n,4,[1,1,1])$ with $n \in \{39, 51\}$ are established via computer search:

i) The 741 codewords of an optimal $(39,4,[1,1,1])_4$-code are given by the cyclic shifts of

$\langle 0,5,28 \rangle$ $\langle 0,19,38 \rangle$ $\langle 0,9,7 \rangle$ $\langle 0,11,20 \rangle$
$\langle 0,35,10 \rangle$ $\langle 0,26,16 \rangle$ $\langle 0,37,13 \rangle$ $\langle 0,14,17 \rangle$
$\langle 0,18,36 \rangle$ $\langle 0,15,21 \rangle$ $\langle 0,6,2 \rangle$ $\langle 0,22,33 \rangle$
$\langle 0,8,30 \rangle$ $\langle 0,1,27 \rangle$ $\langle 0,3,4 \rangle$ $\langle 0,32,25 \rangle$
$\langle 0,12,24 \rangle$ $\langle 0,23,31 \rangle$ $\langle 0,29,34 \rangle$.

ii) The 1275 codewords of an optimal $(51,4,[1,1,1])_4$-code are given by the cyclic shifts of

$\langle 0,36,23 \rangle$ $\langle 0,1,50 \rangle$ $\langle 0,37,11 \rangle$ $\langle 0,12,21 \rangle$
$\langle 0,3,43 \rangle$ $\langle 0,25,2 \rangle$ $\langle 0,8,44 \rangle$ $\langle 0,22,29 \rangle$
$\langle 0,32,16 \rangle$ $\langle 0,17,47 \rangle$ $\langle 0,30,20 \rangle$ $\langle 0,33,14 \rangle$
$\langle 0,49,6 \rangle$ $\langle 0,38,39 \rangle$ $\langle 0,4,26 \rangle$ $\langle 0,45,18 \rangle$
$\langle 0,28,10 \rangle$ $\langle 0,35,15 \rangle$ $\langle 0,7,19 \rangle$ $\langle 0,9,13 \rangle$
$\langle 0,5,42 \rangle$ $\langle 0,31,48 \rangle$ $\langle 0,24,27 \rangle$ $\langle 0,40,34 \rangle$
$\langle 0,41,46 \rangle$.

Next, we give a construction, similar to Theorem 7.3, for optimal $(n,4,[1,1,1])_4$-codes, $n \equiv 3 \pmod{4}$, from $\{3,4\}$-GDDs.

*Theorem 7.5:* Suppose there exists a $\{3,4\}$-GDD of type $9^t u^1, u \in \{0,1,3,4,6,7,9,12,15\}$. Then $A_4(n,4,[1,1,1]) = U(n,4,[1,1,1])$, where $n = 4(9t+u) + 19$.

*Proof:* Let $(X, \mathcal{G}, \mathcal{A})$ be a $\{3,4\}$-GDD of type $9^t u^1, u \in \{0,1,3,4,6,7,9,12,15\}$, with $x$ blocks of size three and $y$ blocks of size four. Then $x$ and $y$ satisfy

$$3x + 6y = \binom{9t+u}{2} - t\binom{9}{2} - \binom{u}{2}. \qquad (3)$$

Now inflate $(X, \mathcal{G}, \mathcal{A})$ by four using a $[1,1,1]$-GDC(4) of type $4^3$ and size 48 (which exists by Theorem 7.1), and a $[1,1,1]$-GDC(4) of type $4^4$ and size 96 (which exists by Proposition 4.5) as ingredients. This gives a $[1,1,1]$-GDC(4) of type $36^t(4u)^1$ having size $48x + 96y$. Now adjoin 19 points to this GDC using

i) a $(4u+19, 4, [1,1,1])_4$-code of size $(4u+19)(2u+9), u \in \{0,1,3,4,6,7,9,12,15\}$,
ii) a $[1,1,1]$-GDC(4) of type $1^{36}19^1$ and size 1314 (which exists by Corollary 7.2),

as ingredients to obtain a $(4(9t+u)+19, 4, [1,1,1])_4$-code of size

$$48x + 96y + 1341t + (4u+19)(2u+9)$$

which simplifies to $\binom{4(9t+u)+19}{2} = U(4(9t+u)+19, 4, [1,1,1])$ using (3).

*Theorem 7.6:* Suppose there exists a $\{3,4\}$-GDD of type $5^t u^1$, and $A_4(4u+11, 4, [1,1,1]) = U(4u+11, 4, [1,1,1])$. Then $A_4(n,4,[1,1,1]) = U(n,4,[1,1,1])$, where $n = 20t + 4u + 11$.

*Proof:* Let $(X, \mathcal{G}, \mathcal{A})$ be a $\{3,4\}$-GDD of type $5^t u^1$, with $x$ blocks of size three and $y$ blocks of size four. Then $x$ and $y$ satisfy

$$3x + 6y = \binom{5t+u}{2} - t\binom{5}{2} - \binom{u}{2}. \qquad (4)$$

Now inflate $(X, \mathcal{G}, \mathcal{A})$ by four using a $[1,1,1]$-GDC(4) of type $4^3$ and size 48 (which exists by Theorem 7.1), and a $[1,1,1]$-

GDC(4) of type $4^4$ and size 96 (which exists by Proposition 4.5) as ingredients. This gives a [1,1,1]-GDC(4) of type $20^t(4u)^1$ having size $48x+96y$. Now adjoin 11 points to this GDC using
 i) a $(4u+11, 4, [1,1,1])_4$-code of size $(4u+11)(2u+5)$,
 ii) a [1,1,1]-GDC(4) of type $1^{20}11^1$ and size 410 (which exists by Proposition 4.8),
as ingredients to obtain a $(20t+4u+11, 4, [1,1,1])_4$-code of size

$$48x + 96y + 410t + (4u+11)(2u+5)$$

which simplifies to $\binom{20t+4u+11}{2} = U(20t+4u+11, 4, [1,1,1])$, using (4). □

To apply Theorems 7.5 and 7.6, we require classes of $\{3,4\}$-GDDs provided below.

*Proposition 7.5:* There exists a $\{3,4\}$-GDD of the following types:
 i) $9^{2t+1}u^1$ with $u = 0, 3, 6, 9, 12, 15$ and $t \geq 2$;
 ii) $9^s u^1$ with $u = 1, 4, 7$ and $s \geq 3$;
 iii) $9^3 u^1$ and $9^4 u^1$ with $u = 0, 3, 6$.

*Proof:*
 i) When $t \geq 2$, take a $\{3\}$-RGDD $(X, \mathcal{G}, \mathcal{A})$ of type $9^{2t+1}$ (which exists by Theorem 3.1), having $\mathcal{A} = \mathcal{A}_1 \cup \cdots \cup \mathcal{A}_{9t}$ as its partition into parallel classes. Now let $\infty_1, \ldots, \infty_u \notin X, u \leq 18$, and consider

$$X' = X \cup \{\infty_1, \ldots, \infty_u\},$$
$$\mathcal{G}' = \mathcal{G} \cup \{\{\infty_1, \ldots, \infty_u\}\},$$
$$\mathcal{A}' = (\cup_{i=1}^{u}\{A \cup \{\infty_i\} : A \in \mathcal{A}_i\})$$
$$\cup (\cup_{i=u+1}^{9t}\{A : A \in \mathcal{A}_i\}).$$

Then $(X', \mathcal{G}', \mathcal{A}')$ is a $\{3,4\}$-GDD of type $9^{2t+1}u^1$.
 ii) For $u = 1, 4, 7$, completing $u$ parallel classes of a 3-RGDD of type $9^3$ gives a $\{3,4\}$-GDD of type $9^3 u^1$. Theorem 3.9 gives 3-GDD's of types $9^{2t}u^1$ with $u = 1, 4, 7$ and $t \geq 2$.
 iii) For $u = 0, 3, 6$, completing $u$ parallel classes of a 3-RGDD of type $9^3$ gives a $\{3,4\}$-GDD of type $9^3 u^1$. Truncating a group of a 4-GDD of type $9^5$ gives $\{3,4\}$-GDD's of types $9^4 u^1$ with $u = 0, 3, 6$. □

*Theorem 7.7:* $A_4(n, 4, [1,1,1]) = U(n, 4, [1,1,1])$ for all $n \equiv 3 \pmod 4, n \geq 39$.

*Proof:* Since the case of $n \equiv 3 \pmod{12}$ is covered by Proposition 7.4, we need only to consider the cases of $n \equiv 7, 11 \pmod{12}$. Proposition 7.5 together with Theorem 7.5 establish the theorem for all $n = 4(9t+u)+19$ with $t \geq 3, u = 0, 3, 6, 1, 4, 7$. This leaves $t = 0, 1, 2$ with $u = 0, 3, 6, 1, 4, 7$ to be considered. These small orders of $n$ are

$$n \equiv 7 \pmod{12} : 19, 31, 43, 55, 67, 79, 91, 103, 115$$

and

$$n \equiv 11 \pmod{12} : 23, 35, 47, 59, 71, 83, 95, 107, 119.$$

Most of them have been constructed previously except for $n \in \{79, 91, 95, 103, 115, 119\}$. For $n \in \{19, 23, 31, 35\}$, we have $A_4(n, 4, [1,1,1]) = U(n, 4, [1,1,1])$ by Proposition 4.4. For $n \in \{43, 47, 59, 67, 71, 83, 107\}$, we have $A_4(n, 4, [1,1,1]) = U(n, 4, [1,1,1])$ by Theorem 7.2. For $n = 55$, we have $A_4(n, 4, [1,1,1]) = U(n, 4, [1,1,1])$ by Proposition 7.2.

For $n = 95$, apply Theorem 7.6 to a $\{3,4\}$-GDD of type $5^3 6^1$ (which exists by Proposition 4.1).

For $n = 119$, we have GDC(4)s of types $6^4$ and $10^4$ by Propositions 4.6 and 4.7. Adjoin an extra point and fill in three of the four groups of a [1,1,1]-GDC(4) of type $10^4$ and size 600 with an $(11, 4, [1,1,1])_4$-code of size 55 to obtain a [1,1,1]-GDC(4) of type $1^{30}11^1$ and size 765. Take a TD(4,5) and truncate it to obtain a $\{3,4\}$-GDD of type $5^3 3^1$. Give weight 6 to obtain a [1,1,1]-GDC(4) of type $30^3 18^1$ and size 3240. Now adjoin 11 extra points and fill in the groups of this GDC with a [1,1,1]-GDC(4) of type $1^{30}11^1$ and size 765 and an optimal $(29, 4, [1,1,1])_4$-code to obtain a $(119, 4, [1,1,1])_4$-code of size 7021, which is optimal.

For the remaining values of $n$, apply Corollary 7.1 to obtain the desired codes noting that $79 = 26 \times 3 + 1, 91 = 30 \times 3 + 1, 103 = 34 \times 3 + 1, 115 = 3 \times 38 + 1$. □

*E. Even Lengths*

Theorems 7.4 and 7.5 combine to show that $A_4(n, 4, [1,1,1]) = U(n, 4, [1,1,1])$ for all odd $n \geq 37$. Chee *et al.* [12, Lemma 21] have proven that if $n$ is odd and $A_4(n, 4, [1,1,1]) = U(n, 4, [1,1,1])$, then $A_4(n-1, 4, [1,1,1]) = U(n-1, 4, [1,1,1])$. Combining with Proposition 4.4, it follows that $A_4(n, 4, [1,1,1]) = U(n, 4, [1,1,1])$ for all even $n \geq 10$, except possibly for $n = 12$.

$A_4(12, 4, [1,1,1]) = U(12, 4, [1,1,1])$ from Corollary 7.1 (noting $A_4(4, 4, [1,1,1]) = U(4, 4, 1[1,1,1])$).

## VIII. SUMMARY

The following summarizes the results obtained in this paper and [12], giving the best state of knowledge about the sizes of optimal $q$-ary constant-composition codes of weight three, for $q > 2$.

*Main Theorem:*
 i) For all integers $n \geq 3$

$$A_3(n, 4, [2,1]) = \begin{cases} \frac{n(n-2)}{4}, & \text{if } n \equiv 0 \pmod 2 \\ \frac{n(n-1)}{4}, & \text{if } n \equiv 1 \pmod 4 \\ \frac{(n-1)^2}{4} + \lfloor \frac{n-3}{12} \rfloor, & \text{if } n \equiv 3 \pmod 4. \end{cases}$$

 ii) For all integers $n \geq 3$

$$A_4(n, 3, [1,1,1]) = \begin{cases} 3, & \text{if } n = 3 \\ n(n-1) - 2, & \text{if } n \in \{5, 6\} \\ n(n-1), & \text{if } n \notin \{3, 5, 6\}. \end{cases}$$

 iii) For all integers $n \geq 3$

$$A_4(n, 4, [1,1,1]) = \begin{cases} 1, & \text{if } n = 3 \\ 6, & \text{if } n = 5 \\ 11, & \text{if } n = 6 \\ 16, & \text{if } n = 7 \\ 23, & \text{if } n = 8 \\ n \lfloor \frac{n-1}{2} \rfloor, & \text{if } n \notin \{3, 5, 6, 7, 8\} \end{cases}$$

except possibly for $n \in \{9, 13, 15, 17\}$.

iv) For all integers $n \geq 3$

$$A_4(n, 5, [1,1,1]) = \begin{cases} 1, & \text{if } n \leq 4 \\ 2, & \text{if } n = 5 \\ 4, & \text{if } n = 6 \\ n, & \text{if } n \geq 7. \end{cases}$$

## IX. CONCLUSION

We introduced the concept of group divisible codes, which share similarities with group divisible designs that allow the use of powerful Wilson-type constructions. This class of codes was shown to play an important role in determining the size of optimal constant-composition codes. In particular, the sizes of optimal constant-composition codes of weight three (and specified distance) are determined, leaving only four outstanding cases.

We are optimistic that group divisible codes will find further applications in the determination of optimal constant-composition codes and constant-weight codes.